\documentclass[12pt,a4paper]{article}

\usepackage{epsfig}
\usepackage{a4wide}

\renewcommand{\theequation}{\thesection.\arabic{equation}}
\newcommand{\cN}{{\mathcal N}}

\newcommand{\ep}{{\epsilon}} 

\newcommand{\MS}{{\mathrm{MS}}}

\makeatletter
\def\mr@ignsp#1 {\ifx\:#1\@empty\else #1\expandafter\mr@ignsp\fi}%
\newcommand{\multiref}[1]{\begingroup
\xdef\mr@no@sparg{\expandafter\mr@ignsp#1 \: }%
\def\mr@comma{}%
\@for\mr@refs:=\mr@no@sparg\do{\mr@comma\def\mr@comma{,}\ref{\mr@refs}}%
\endgroup}
\makeatother

\def\pqqx{p_{\rm{qq}}(x)}
\def\pqqmx{p_{\rm{qq}}(-x)}
\def\HP{{\mathrm{H}}}
\def\HS{S}
\def\z#1{{\zeta_{#1}}}
\def\ca{{C^{}_A}}
\def\cf{{C^{}_F}}
\def\nf{{n^{}_{\! f}}}
\def\n2f{{n^{\,2}_{\! f}}}
\def\TF{{T^{}_{\! F}}}
\def\T2F{{T^{\,2}_{\! F}}}
\def\S(#1){{{S}_{#1}}}
\def\Ss(#1,#2){{{S}_{#1,#2}}}
\def\N0{{{1}}}
\def\Npm{{{\bf N_{\pm}}}}
\def\Npmi{{{\bf N_{\pm i}}}}
\def\Nminus{{{\bf N_{-}}}}
\def\Nplus{{{\bf N_{+}}}}
\def\Nminustwo{{{\bf N_{-2}}}}
\def\Nplustwo{{{\bf N_{+2}}}}
\def\pqq(#1){p_{\rm{qq}}(#1)}
\def\H(#1){{\rm{H}}_{#1}}
\def\Hh(#1,#2){{\rm{H}}_{#1,#2}}
\def\Hhh(#1,#2,#3){{\rm{H}}_{#1,#2,#3}}
\def\Hhhh(#1,#2,#3,#4){{\rm{H}}_{#1,#2,#3,#4}}
\def\M{N}
\newcommand{\bra}[1]{\langle#1\mid}
\newcommand{\ket}[1]{\mid#1\rangle}

\begin{document}
\thispagestyle{empty}

\begin{center}
{\Large{\bf
Three-loop anomalous dimension\\[3mm] of the non-singlet transversity operator in QCD
}}
\vspace{15mm}

{\sc
V.~N.~Velizhanin}\\[5mm]

{\it Theoretical Physics Department\\
Petersburg Nuclear Physics Institute\\
Orlova Roscha, Gatchina\\
188300 St.~Petersburg, Russia}\\[5mm]

\textbf{Abstract}\\[2mm]
\end{center}

\noindent{We calculate the three-loop anomalous dimension of the non-singlet transverse operator from $N=1$ to $N=15$. Using some guess we have reconstructed a general form of three-loop anomalous dimension for arbitrary Mellin moment $N$.
Obtained result is transformed into Bjorken-$x$ space by an inverse Mellin transformation.
The final expressions are presented in both Mellin-$N$ and Bjorken-$x$ space.
}
\newpage

\setcounter{page}{1}
\section{Introduction}
\label{sec:intro}

The anomalous dimensions of the non-singlet Wilson twist-2 operator are known up to three-loop order
\cite{Gross:1973ju,Georgi:1951sr,
Altarelli:1977zs,
Floratos:1977au,
GonzalezArroyo:1979df,
Curci:1980uw,
Moch:2004pa}
and the first even moment was calculated at four loops \cite{Baikov:2006ai,Velizhanin:2011es}. There is a similar non-singlet transversity operator, which appear, for example, in a study of semi-inclusive deeply-inelastic scattering (SIDIS)
\cite{Ralston:1979ys,
Jaffe:1991kp,
Jaffe:1991ra,
Cortes:1991ja}
and in the polarized Drell-Yan process
\cite{Cortes:1991ja,
Collins:1992kk,
Jaffe:1993xb,
Tangerman:1994bb,
Boer:1997nt,
Artru:1989zv}.
The scaling violations of the transversity distribution were explored in leading
\cite{Artru:1989zv,
Baldracchini:1980uq,
Shifman:1980dk,
Bukhvostov:1985rn,
Blumlein:2001ca,
Mukherjee:2001zx}
and next-to-leading order \cite{Hayashigaki:1997dn,Kumano:1997qp,Vogelsang:1997ak}.
At three-loop order the moments~$N = 1$ to~$8$ for the anomalous dimension are known~\cite{Gracey:2003yr,Gracey:2003mr,Gracey:2006zr,Gracey:2006ah} and for~$N~=~9$~to~$14$ there are results for~${\mathcal {O}}(\nf\TF)$ contributions~\cite{Blumlein:2009rg}.
In spite of the calculations of three-loop anomalous dimension for this operator do not seem very complicated with compare to already performed in Refs.~\cite{Moch:2004pa,Vogt:2004mw,Vermaseren:2005qc,Moch:2008fj} it still absent.
In this paper, however, we want to show how it is possible to obtain such result using the methods, which we developed for the computations of the anomalous dimensions for composite operators in the maximally extended $\cN=4$ Supersymmetric Yang-Mills (SYM) theory~\cite{Kotikov:2003fb,Kotikov:2004er,Kotikov:2007cy,Lukowski:2009ce,Velizhanin:2010cm}. Namely, using presented below the first fifteen Mellin moments and some guess we have reconstructed a general form of three-loop anomalous dimension for the arbitrary Mellin moment $N$. In spite of the fact that the obtained result seems to be correct we should stress, that it is mathematically non-rigorous. More consequent method for such calculations was suggested in Ref.~\cite{Blumlein:2009tj}. Nevertheless, the result of the recent calculations of the next ($N=16$) Mellin moment of the three-loop anomalous dimension of the non-singlet transversity operator from Ref.~\cite{Bagaev:2012bw} coincides with the prediction, coming from our result and can serve as a direct confirmation of the general result presented in this paper, at least for even values of $N$.

The article is organized as follows: In section \ref{sec:Calculation} we briefly describe the method,
which we used for calculation of fixed moments of anomalous dimension of the non-singlet transversity operator.
The reconstruction of the three-loop non-singlet transversity anomalous dimension from the calculated values is performed
in section~\ref{sec:Reconstruction}.
The methods of reconstruction are described in sections~\ref{subsec:cfn2f} and~\ref{subsec:cf3}.
The final result is presented  in section~\ref{sec:ResultsN} in Mellin-$N$ space.
In section~\ref{sec:ResultsX} we present our result for $x$-space splitting functions. Finally we briefly summarize our findings in section~\ref{sec:Conclusion}.
In Appendix we write down the results for fixed moments of anomalous dimension of non-singlet transversity~operator~from~$N = 1$~to~$15$.

\setcounter{equation}{0}
\section{Calculation}\label{sec:Calculation}

Our calculations of the fixed moments of anomalous dimension for the non-singlet transversity operator were performed in the same way, as in~\cite{Blumlein:2009rg} except the last step, where we used MINCER~\cite{Gorishnii:1989gt} instead of MATAD~\cite{Steinhauser:2000ry}, as we interesting only with the divergences. We repeat here main futures from~\cite{Blumlein:2009rg} and detailed description can be found in original~paper.

The local flavor non-singlet twist-2 transversity operator is given by
\begin{equation}
\label{op2}
{\mathcal O}_{q,r}^{{\mathrm {TR,ns}}, \mu, \mu_1, \ldots, \mu_N}(z) =
\frac{1}{2} i^{N-1} S
\left[\overline{q}(z) \sigma^{\mu \mu_1} {\mathcal D}^{\mu_2} \ldots {\mathcal D}^{\mu_N} \frac{\lambda_r}{2} q(z)\right],
\end{equation}
with $\sigma^{\mu\nu} = (i/2)\left[\gamma^\mu \gamma^\nu - \gamma^\nu \gamma^\mu \right]$, $\lambda_r$ the Gell-Mann matrices for $SU(3)_{\rm flavor}$, ${\mathcal D}^{\mu}$ the covariant derivative in QCD, $q (\overline{q})$ denote the quark and antiquark fields, and the operator $S$ symmetrizes the Lorentz indices and subtracts the trace terms.

Following Ref.~\cite{Bierenbaum:2009mv} we consider the Green's function $\hat{G}^{ij,{\mathrm {TR,ns}}}_{\mu,q}$ which is obtained by contracting the matrix element of local operator (\ref{op2}) with the source term $J_N = \Delta^{\mu_1} \ldots \Delta^{\mu_N}$
\begin{equation}
\label{GijTRNS}
\overline{u}(p,s) G^{ij, {\mathrm {TR,ns}}}_{\mu,q} \lambda_r u(p,s)=
J_N\bra{q_i(p)}{\mathcal O}_{q,r;\mu, \mu_1, \ldots, \mu_N}^{{\mathrm {TR,ns}}}\ket{q^j(p)}\,,
\end{equation}
where $p$ and $s$ denote the four-vectors of the momentum and spin of the external quark line, $u(p,s)$ is the corresponding bi-spinor, $\Delta^2 = 0$.
The contraction with the source term $J_N = \Delta^{\mu_1}\ldots \Delta^{\mu_N}$ allows to write a general expression for the corresponding projector, which can be found in Ref.~\cite{Bierenbaum:2009mv}.

The unrenormalized Green's function has the following Lorentz structure~\cite{Blumlein:2009rg}
\begin{eqnarray}
\hspace{-8mm}\hat{G}^{ij, {\mathrm {TR,ns}}}_{\mu,q}&\!=&\!
\delta_{ij}(\Delta \cdot p)^{N-1}\!
\Big(
\Delta_{\rho}\sigma^{\mu\rho}
\,\Sigma^{\mathrm {TR,ns}}(p)
+\!c_1 \Delta^\mu 
+\! c_2 p^\mu
+\!c_3 \gamma^\mu p \hspace*{-2mm} /
+\!c_4 \Delta \hspace*{-3mm}/~p \hspace*{-2mm}/ \Delta^\mu
+\!c_5 \Delta \hspace*{-3mm}/~p \hspace*{-2mm}/ p^\mu
\Big)\!,
\end{eqnarray}
with unphysical constants $c_k|_{k=1 ... 5}$.
To determine $\Sigma^{\mathrm {TR,ns}}(p)$ we use, following Ref.~\cite{Blumlein:2009rg}, the following projector ($N_c$ denotes the number of colors)
\begin{eqnarray}
\label{eqc3}
\Sigma^{\mathrm {TR,ns}}(p) &=&
- i \frac{\delta^{ij}}{4N_c(\Delta.p)^{N+1} (D-2)}
\Big\{
{\mathrm {Tr}}[ \Delta\hspace*{-3mm}/~p\hspace*{-2mm}/~
p^{\mu}\hat{G}^{ij, {\mathrm {TR,ns}}}_{\mu,q}]
-\Delta.p {\mathrm {Tr}}[p^{\mu}\hat{G}^{ij, {\mathrm {TR,ns}}}_{\mu,q}]
\nonumber\\ && \hspace{50mm}
+i\Delta.p {\mathrm {Tr}}[\sigma_{\mu \rho} p^\rho
\hat{G}^{ij, {\mathrm {TR,ns}}}_{\mu,q}]
\Big\}\,.
\end{eqnarray}

Renormalization constants within $\MS$-like schemes do not depend on dimensional parameters (masses, momenta)~\cite{Collins:1974bg} and have the following structure
\begin{equation}
\label{eq:5}
Z_\Gamma\!\left(\frac{1}{\epsilon},\alpha,g^2\right)=
1+\sum^\infty_{n=1}c_\Gamma^{(n)}\!\left(\alpha,g^2\right)\epsilon^{-n},
\end{equation}
where $\alpha$ is the gauge fixing parameter.
The renormalization constants define corresponding anomalous dimensions:
\begin{equation}
\label{defga}
\gamma_\Gamma(\alpha,g^2)=
g^2\frac{\partial}{\partial g^2}\ c^{(1)}_\Gamma(\alpha,g^2)=\sum^\infty_{n=1}\gamma_\Gamma^{(n-1)}g^{2n}.
\end{equation}

For the calculation of the renormalization constants, following Ref.~\cite{Larin:1993tp} (see also Refs.~\cite{Tarasov:1976ef,Vladimirov:1979zm,Tarasov:1980kx}), we use the multiplicative renormalizability of Green's functions.
The renormalization constants $Z_\Gamma$ relate the dimensionally regularized one-particle-irreducible Green's functions with renormalized one as
\begin{equation}
\label{multren}
\Gamma_{\mathrm{Renormalized}}\left(\frac{Q^2}{\mu^2},\alpha,g^2\right)=\lim_{\epsilon \rightarrow 0}
Z_\Gamma\left(\frac{1}{\epsilon},\alpha,g^2\right)
\Gamma_{\mathrm{Bare}}\left(Q^2,\alpha_{\mathrm{B}},g^2_\mathrm{B},\epsilon\right),
\end{equation}
where $g^2_{\mathrm{B}}$ and $\alpha_{\mathrm{B}}$ are the bare charge and the bare gauge fixing parameter 
with
\begin{equation}\label{gbex}
g^2_{\mathrm{B}}=
\mu^{2\epsilon}\left[g^2+\sum_{n=1}^{\infty}a^{(n)}\!\left(g^2\right)\epsilon^{-n}\right], 
\qquad\alpha_{\mathrm{B}}=\alpha Z_3\,,\qquad Z_3=Z_g\,.
\end{equation}
To obtain the anomalous dimension of transversity operator we should subtract renormalization of the external quark field, so
\begin{equation}
Z_{\mathcal{O}^{\,\mathrm {TR,ns}}}=Z_{q}^{-1}\,Z_{\Sigma^{\,\mathrm {TR,ns}}}\,,\label{gbz}
\end{equation}
where $Z_q$ is the renormalization constant for the quark field.

To find the renormalization constant we compute with the FORM~\cite{Vermaseren:2000nd} package MINCER~\cite{Gorishnii:1989gt} the unrenormalized three-loop
$\Sigma^{\mathrm {TR,ns}}(p)$ and determine its renormalization constant $Z_{\Sigma^{\mathrm {TR,ns}}}$ from the requirement that the poles in $\ep$ cancel in the right-hand side of Eq.~(\ref{multren}).
As in our previous calculations of the three-loop renormalization constants in supersymmetric Yang-Mills theories~\cite{Velizhanin:2008rw}
we used a program DIANA~\cite{Tentyukov:1999is}, which call QGRAF~\cite{Nogueira:1991ex} to generate all diagrams and the FORM package COLOR~\cite{vanRitbergen:1998pn} for evaluation of the color traces.
Because the right-hand side of Eq.~(\ref{multren}) contains the bare gauge fixing parameter $\alpha_{\mathrm{B}}$, we should perform all calculations up to two loops with the arbitrary gauge fixing parameter~$\alpha$ (i.e. the propagator of gluon is $(g_{\mu\nu}-(1-\alpha)q_\mu q_\nu/q^2)/q^2$), while for the three-loop calculations we used Feynman gauge $\alpha=1$. To obtain result for the renormalization constant from Eq.~(\ref{multren}), we put $\alpha=1$ only after expansion in the right-hand side of Eq.~(\ref{multren}).
The anomalous dimensions $\gamma^{\,\pm}_{\,\mathrm {TR,ns}}(N)$ are expanded in powers of $\alpha_s /(4\pi)$ as
\begin{equation}\label{expAD}
\gamma^{\,\pm}_{\,\mathrm {TR,ns}}(N)=
\frac{\alpha_s}{4\pi}
\ \gamma^{(0)\pm}_{\,\mathrm {TR,ns}}(N)
+
\left(\frac{\alpha_s}{4\pi}\right)^2
\!\gamma^{(1)\pm}_{\,\mathrm {TR,ns}}(N)
+
\left(\frac{\alpha_s}{4\pi}\right)^3
\!\gamma^{(2)\pm}_{\,\mathrm {TR,ns}}(N)
+\ldots\,.
\end{equation}
We have calculated the fifteen first moments of the three-loop anomalous dimension $\gamma^{(2)\pm}_{\,\mathrm {TR,ns}}$ of the non-singlet transversity operator~(\ref{op2}), which are listed in Appendix.

\setcounter{equation}{0}
\section{Reconstruction}\label{sec:Reconstruction}

Having in hand fifteen first values for the three-loop anomalous dimension of the flavour non-singlet transversity operator we have tried to reconstruct a general form of anomalous dimension equipped with our experience in such exercise, which we have done during the computations of higher orders anomalous dimensions of composite operators in the maximally extended \mbox{$\cN=4$}~SYM theory~\cite{Kotikov:2003fb,Kotikov:2004er,Kotikov:2007cy,Lukowski:2009ce,Velizhanin:2010cm}.

As well known from the direct calculations in QCD the anomalous dimensions of \mbox{twist-2} operators are expressed in terms of harmonic sums~\cite{GonzalezArroyo:1979df,GonzalezArroyo:1979he,Vermaseren:1998uu,Blumlein:1998if}, following Ref.~\cite{Vermaseren:1998uu} recursively defined by
\begin{equation}\label{vhs}
S_{\pm m} (\M)=\sum^{\M}_{i=1} \frac{(\pm 1)^{i}}{i^{m}}\, , \qquad
S_{\pm m_1,m_2,\ldots,m_k}(\M)=\sum^{\M}_{i=1} \frac{(\pm 1)^{i}}{i^{m_1}}
\,S_{m_2,\ldots,m_k}(i)\, .
\end{equation}
The sum of the absolute values of the indices $m_i$ defines the weight of the harmonic sum. In the $n$-loop anomalous dimensions one can encounter sums up to weight $2n-1$.
Thus, the harmonic sums form the basis and if we calculate a sufficient number of fixed moments of the anomalous dimension we can restore a general form of anomalous dimension. We used such method earlier for the computations of anomalous dimensions at two-loop~\cite{Kotikov:2003fb}, three-loop~\cite{Kotikov:2004er}, four-loop~\cite{Kotikov:2007cy} and five-loop~\cite{Lukowski:2009ce} orders for the twist-2 operators and even at six-loop~\cite{Velizhanin:2010cm} order for the twist-3 operator in $\cN=4$ SYM theory.

To find a minimal basis we used the maximal transcendentality principle, which was discovered during the calculations of the eigenvalue of the kernel of Balitsky-Fadin-Kuraev-Lipatov (BFKL) equation~\cite{Lipatov:1976zz,Kuraev:1977fs,Balitsky:1978ic} at the next-to-leading logarithm approximation in $\cN=4$ SYM theory~\cite{Kotikov:2002ab}. It was found, that the final result contains only the most complicated functions from the full result in QCD~\cite{Fadin:1998py}. Assuming a deep relation between Dokshitzer-Gribov-Lipatov-Altarelli-Parisi (DGLAP)~\cite{Gribov:1972ri,Gribov:1972rt,Altarelli:1977zs,Dokshitzer:1977sg} and BFKL equations, the authors of Ref.~\cite{Kotikov:2002ab} suggested that the eigenvalues of the anomalous dimension matrix of twist-2 operators in $\cN=4$ SYM theory should contain the most complicated harmonic sums, i.e. the harmonic sums only with the maximal weight (maximal transcendentality). This hypothesis was confirmed by direct calculation at the two loops in Ref.~\cite{Kotikov:2003fb}.
So, the result in the maximally extended \mbox{$\cN=4$}~SYM theory can be obtained from the corresponding result in QCD, if we take from the QCD result only those terms, which have the harmonic sums with maximal transcendentality (with maximal weight).
Note, that the eigenvalues of the anomalous dimension matrix of twist-2 operators in supersymmetric theories can be expressed through one function with shifted argument. In Ref.~\cite{Kotikov:2002ab}  such function was called an universal anomalous dimension. The universal anomalous dimension of the $\cN=4$ SYM theory is contained in any diagonal element of the anomalous dimension matrix of twist-2 operators in any four-dimensional gauge theory, for example, in QCD, but with corresponding colour factors.

The anomalous dimension of the non-singlet transversity operator should contain the same universal anomalous dimension as a part of full result. Moreover, one can see, that one- and two-loop anomalous dimensions of the non-singlet transversity operator~\cite{Baldracchini:1980uq,Shifman:1980dk,Bukhvostov:1985rn,Artru:1989zv,Blumlein:2001ca,Hayashigaki:1997dn,Kumano:1997qp,Vogelsang:1997ak}
 differ only by few terms with compare to the universal anomalous dimension~\cite{Kotikov:2002ab,Kotikov:2003fb}. We assume, that this property will be hold also at three loops.

Let's see how it works at two loops. We take the result for the two-loop anomalous dimension of the non-singlet operator from Eq.~(3.5) of Ref.~\cite{Moch:2004pa}
\begin{eqnarray}
  &&\gamma^{\,(1)+}_{\,\rm ns}(\M) =
 4\, \* {\cf \* \nf} \* \bigg(
            {1 \over 12}
          + {4 \over 3} \* \S(1)
          - (\Nminus+\Nplus) \* \bigg[
            {11 \over 9} \* \S(1)
          - {1 \over 3} \* \S(2)
          \bigg]
          \bigg)
+ 4\, \* {\cf^{2}} \* \bigg(
            4 \* \S(-3)
          + 2 \* \S(1)
  \nonumber\\[3mm]&&\quad
          + 2 \* \S(2)
          - {3 \over 8}
          + \Nminus \* \bigg[
            \S(2)
          + 2 \* \S(3)
          \bigg]
          - (\Nminus+\Nplus) \* \bigg[
            \S(1)
          + 4 \* \Ss(1,-2)
          + 2 \* \Ss(1,2)
          + 2 \* \Ss(2,1)
          + \S(3)
          \bigg]
          \bigg)
  \nonumber\\[3mm]&&\quad
 + 4\, \* {\ca \* \cf} \* \bigg(
            2\, \* \Nplus \* \S(3)
          - {17 \over 24}
          - 2 \* \S(-3)
          - {28 \over 3} \* \S(1)
          + (\Nminus+\Nplus) \* \bigg[
            {151 \over 18} \* \S(1)
          + 2 \* \Ss(1,-2)
          - {11 \over 6} \* \S(2)
          \bigg]
          \bigg),\qquad\quad\label{eq:gqq1p}
 \\[4mm]
  &&\gamma^{\,(1)-}_{\,\rm ns}(\M) =
     \gamma^{\,(1)+}_{\,\rm ns}(\M)
\nonumber\\[3mm]&& \qquad\qquad\qquad
      + 16\, \*  {\cf \* \bigg(\cf - {\ca \over 2} \bigg)}  \*  \bigg(
            (\Nminus-\Nplus) \* \bigg[
            \S(2)
          - \S(3)
          \bigg]
          - 2 \* (\Nminus+\Nplus-2) \* \S(1)
          \bigg).
 \label{eq:gqq1m}
\end{eqnarray}
where $  \Npm \, S_{\vec{m}} = S_{\vec{m}}(N \pm 1)$ and make substitutions $\Nminus\to 1$ and $\Nplus\to 1$. As result, we obtain
\begin{eqnarray}
  &&\widehat\gamma^{\,(1)+}_{\,\rm ns}(\M) =
4\, \* {\cf \* \nf} \* \bigg(
            {1 \over 12}
          - {10 \over 9} \* \S(1)
          + {2 \over 3} \* \S(2)
          \bigg)
+ 4\, \* {\cf^{2}} \* \bigg(
            4 \* \S(-3)
          + 3 \* \S(2)
          - {3 \over 8}
          + 2 \* \S(3)
          - 8 \* \Ss(1,-2)
  \nonumber\\&& \mbox{} \quad
          - 4 \* \Ss(1,2)
          - 4 \* \Ss(2,1)
          - 2 \S(3)
          \bigg)
+  4\, \* {\ca \* \cf} \* \bigg(
            2\, \* \S(3)
          - {17 \over 24}
          - 2 \* \S(-3)
          + {67 \over 9} \* \S(1)
          + 4 \* \Ss(1,-2)
          - {11 \over 3} \* \S(2)
          \bigg),\qquad\ \label{eq:gqqHat1p}
 \\[2mm]
  &&\widehat\gamma^{\,(1)-}_{\,\rm ns}(\M) =
     \widehat\gamma^{\,(1)+}_{\,\rm  ns}(\M),
 \label{eq:gqqHat1m}
\end{eqnarray}
which should be compared with the two-loop anomalous dimension of transversity operator, calculated in Refs.~\cite{Hayashigaki:1997dn,Kumano:1997qp,Vogelsang:1997ak}:
\begin{eqnarray}
&&\gamma^{(1)}_{\rm TR,ns}(\M) =
4\, \* {\cf \* \nf} \* \bigg(
            {1 \over 12}
          - {10 \over 9} \* \S(1)
          + {2 \over 3} \* \S(2)
          \bigg)
+ 4\, \* {\cf^{2}} \* \bigg(
            4 \* \S(-3)
          + 3 \* \S(2)
          - {3 \over 8}
          + 2 \* \S(3)
          - 8 \* \Ss(1,-2)
  \nonumber\\&& \mbox{} \quad
          - 4 \* \Ss(1,2)
          - 4 \* \Ss(2,1)
          - 2 \S(3)
          \bigg)
+  4\, \* {\ca \* \cf} \* \bigg(
            2\, \* \S(3)
          - {17 \over 24}
          - 2 \* \S(-3)
          + {67 \over 9} \* \S(1)
          + 4 \* \Ss(1,-2)
          - {11 \over 3} \* \S(2)
          \bigg)
\nonumber\\[2mm]
&& \mbox{} \quad- 8 C_F\left(C_F - \frac{C_A}{2}\right)\frac{1 -(-1)^{N}}{N(N+1)}
\ .\label{eq:gqqTR1p}
\end{eqnarray}
The difference is only in the last term.
So, we can expect, that such simplification will work at the three-loop order too.

\subsection{$\cf\,\n2f$ contribution}\label{subsec:cfn2f}

We start the reconstruction of the three-loop anomalous dimension of non-singlet transversity operator from the most simple contribution proportional to $\cf\n2f$, which already known from previous calculations~\cite{Gracey:2003mr,Ablinger:2010ty}.
Again, we take Eq.~(3.7) from arXiv version of Ref.~\cite{Moch:2004pa}
\begin{equation}\label{cfn2fMVV}
\gamma_{\mathrm{ns},\cf\n2f}^{(2)+}(\M)= 16\,{\cf  \n2f}  \bigg(
            {17 \over 144}
          - {13 \over 27}  \S(1)
          + {2 \over 9}  \S(2)
              + (\Nminus+\Nplus)   \bigg[
            {2 \over 9}  \S(1)
          - {11 \over 54}  \S(2)
          + {1 \over 18}  \S(3)
          \bigg]
          \bigg)\,.
\end{equation}
In this equation we make substitutions $\Nminus\to 1$ and $\Nplus\to 1$ and obtain
\begin{equation}\label{gammahatcfnf2}
\widehat{\gamma}_{\mathrm{ns},\cf\n2f}^{\,(2)}(\M)= 16\,{\cf  \n2f}  \bigg(
            {17 \over 144}
          - {1 \over 27}  \S(1)
          - {5 \over 27}  \S(2)
          + {1 \over 9}  \S(3)
          \bigg)\,.
\end{equation}
Then we subtract the last equation at $\M=1,\ldots,15$ from calculated values Eqs.~(\ref{res1})-(\ref{res15}) with the following result
\begin{equation}\label{cfnf2red}
-\cf\n2f\!\left\{
\frac{4}{9}\,,
\frac{4}{27}\,,
\frac{2}{27}\,,
\frac{2}{45}\,,
\frac{4}{135}\,,
\frac{4}{189}\,,
\frac{1}{63}\,,
\frac{1}{81}\,,
\frac{4}{405}\,,
\frac{4}{495}\,,
\frac{2}{297}\,,
\frac{2}{351}\,,
\frac{4}{819}\,,
\frac{4}{945}\,,
\frac{1}{270}
\right\}\!.
\end{equation}

To reconstruct a full $N$ dependence for obtained numbers we should write a suitable basis from the harmonic sums and solve the system of equations to find coefficients in the front of sums in ansatz. Assumption about the minimal basis can be obtained with the help of \texttt{FactorInteger} function from \texttt{MATHEMATICA} applied to the denominators. Such analysis allows to suggest the following basis:
\begin{equation}\label{Basisnf2}
\mathcal{B}_{\cf\n2f}=
\left\{
\HS_1\,,\
\frac{1}{\M}\,,\
\frac{1}{\M+1}\,,\
1
\right\}
\end{equation}
and it is easy to find the general expression for Eq.~(\ref{cfnf2red}):
\begin{equation}\label{cfnf2redM}
\widetilde{\gamma}_{\mathrm{TR,ns},\,\cf\n2f}^{\,(2)}(\M)=
-\frac{8}{9}\,\frac{\cf\n2f}{\M(\M+1)}\ .
\end{equation}
So, the contribution, proportional to $\cf\n2f$ is equal to
\begin{equation}\label{cfn2fTR}
\gamma_{\mathrm{TR},\mathrm{ns},\cf\n2f}^{(2)}(\M)= 16\,{\cf  \n2f}  \bigg(
            {17 \over 144}
          - {1 \over 27}\,  \S(1)
          - {5 \over 27}\,  \S(2)
          + {1 \over 9} \, \S(3)
-\frac{1}{18}\,\frac{1}{\M(\M+1)}
          \bigg)\,,
\end{equation}
which is a full agreement with known result~\cite{Gracey:2003mr,Ablinger:2010ty}.
Note, that constant is the same both for non-singlet anomalous dimension~(\ref{cfn2fMVV}) and for non-singlet transversity anomalous dimension~(\ref{cfn2fTR}).

\subsection{$\cf^2\nf$ contribution}\label{subsec:cf2nf}

Now we proceed with the same method to reconstruct the rest parts.
Again, we substitute $\Nminus\to 1$ and $\Nplus\to 1$ into the result for the three-loop non-singlet anomalous dimension from Ref.~\cite{Moch:2004pa}  and obtain for $\cf^2\nf$ contribution
\begin{eqnarray}
\widehat{\gamma}_{\mathrm{ns},\cf^2\nf}^{\,(2)}(\M)&=& 16\,\cf^2\nf \bigg(
\frac{23}{16}
-\frac{55}{24} \HS_1
-\frac{5}{12} \HS_2
+\HS_3
+\frac{2}{3} \HS_4
-\frac{20}{9} \HS_{-3}
+\frac{4}{3} \HS_{-4}\nonumber\\
&&+\frac{4}{3} \HS_{-3,1}
-\frac{4}{3} \HS_{1,-3}
+\frac{40 }{9}\HS_{1,-2}
+\frac{20}{9} \HS_{1,2}
-\frac{8}{3} \HS_{1,3}
-\frac{4}{3} \HS_{2,-2}\nonumber\\
&&+\frac{20}{9} \HS_{2,1}
-\frac{4}{3} \HS_{2,2}
-\frac{4}{3} \HS_{3,1}
-\frac{8}{3} \HS_{1,-2,1}
+\left(2\,\HS_1-\frac32\right)\z3
\bigg)\,.\label{gammahatcf2nf}
\end{eqnarray}
Subtracting the last equation at $\M=1,\ldots,15$ from calculated values Eqs.~(\ref{res1})-(\ref{res15}) we have found the following result:
\begin{eqnarray}
&\cf^2\nf&\bigg\{
\frac{56}{9}\,,
\frac{4}{3}\,,
\frac{8}{9}\,,
\frac{5}{9}\,,
\frac{206}{675}\,,
\frac{14}{45}\,,
\frac{313}{2205}\,,
\frac{761}{3780}\,,
\frac{3287}{42525}\,,
\frac{671}{4725}\,,
\frac{31789}{686070}\,,\nonumber\\[2mm]&&\hspace*{50mm}
\frac{6617}{62370}\,,
\frac{364087}{12297285}\,,
\frac{1171733}{14189175}\,,
\frac{160379}{8108100}
\bigg\}.\label{cf2nfred}
\end{eqnarray}
Again, analyzing denominators of obtained numbers we can write the following basis:
\begin{equation}\label{Basiscf2nf}
\mathcal{B}_{\cf^2\nf}=\left\{
\frac{1}{\M}\,,\
\frac{1}{\M+1}\,,\
\frac{1}{\M^2}\,,\
\frac{1}{(\M+1)^2}\,,\
\frac{\HS_1}{\M}\,,\
\frac{\HS_1}{\M+1}
\right\}\,.
\end{equation}
With this basis it is easy to find the general expression for Eq.~(\ref{cf2nfred}) for even values of $\M$
\begin{equation}\label{cf2nfredMP}
\widetilde{\gamma}_{\mathrm{TR,ns},\cf^2\nf}^{\,(2)+}(\M)=
\frac{16}{3}\,\cf^2\nf\frac{\HS_1}{\M(\M+1)}
\end{equation}
and for odd values of $\M$
\begin{equation}\label{cf2nfredMM}
\widetilde{\gamma}_{\mathrm{TR,ns},\cf^2\nf}^{\,(2)-}(\M)=\frac{16}{3}\,\cf^2\nf\left(
\frac{\HS_1}{\M(\M+1)}
-\frac{7}{3}\frac{1}{\M(\M+1)}
-\frac{2}{(\M+1)^2}
\right).
\end{equation}

Note, that for $\z3$ part our reduction rule gives correct result immediately
\begin{equation}\label{cf2nfz3redM}
\widehat{\gamma}_{\mathrm{ns},\cf^2\nf,\z3}^{\,(2)}(\M)=
\cf^2 \nf \left(32 \HS_1-24\right) \z3\,.
\end{equation}

The final expression for $\cf^2\nf$ contribution can be found from Eqs.~(\ref{gammahatcf2nf}), (\ref{cf2nfredMP}) and (\ref{cf2nfredMM}):
\begin{equation}\label{cf2nfTR}
{\gamma}_{\mathrm{TR,ns},\cf^2\nf}^{\,(2)\pm}(\M)=
\widehat{\gamma}_{\mathrm{ns},\cf^2\nf}^{\,(2)}(\M)+
\widetilde{\gamma}_{\mathrm{TR,ns},\cf^2\nf}^{\,(2)\pm}(\M)\,.
\end{equation}

\subsection{$\ca\cf\nf$ contribution}\label{subsec:cacfnf}

For $\ca\cf\nf$ contribution our reduction rule $\Nminus\to 1$ and $\Nplus\to 1$ for the three-loop non-singlet anomalous dimension from Ref.~\cite{Moch:2004pa}  gives
\begin{eqnarray}
\widehat{\gamma}_{\mathrm{ns},\ca\cf\nf}^{\,(2)}(\M)&=& 16\,\ca\cf \nf \bigg(
-\frac54
-\frac{209}{108} \HS_1
+\frac{167}{54} \HS_2
-\frac{7}{3} \HS_3
+\frac{2}{3} \HS_4
+\frac{10}{9} \HS_{-3}
\nonumber\\&&
-\frac{2}{3} \HS_{-4}
+ \HS_{1,3}
-\frac{2}{3} \HS_{-3,1}
+\frac{2}{3} \HS_{1,-3}
-\frac{20}{9} \HS_{1,-2}
+\frac{2}{3} \HS_{2,-2}\nonumber\\
&&-\frac{1}{3} \HS_{3,1}
+\frac{4}{3}\HS_{1,-2,1}
-\left(2\,\HS_1-\frac32\right)\z3
\bigg)\label{gammahatcacfnf}
\end{eqnarray}
and subtracting the last equation at $\M=1,\ldots,15$ from calculated values Eqs.~(\ref{res1})-(\ref{res15}) we obtain
\begin{eqnarray}
&\ca\cf\nf&\bigg\{
\frac{22}{9}\,,
\frac{35}{27}\,,
\frac{31}{54}\,,
\frac{7}{20}\,,
\frac{371}{1350}\,,
\frac{293}{1890}\,,
\frac{2917}{17640}\,,
\frac{3877}{45360}\,,
\frac{47}{420}\,,
\frac{1009}{18900}\,,\nonumber\\[2mm]&&\hspace*{30mm}
\frac{74471}{914760}\,,
\frac{117259}{3243240}\,,
\frac{340051}{5465460}\,,
\frac{1470907}{56756700}\,,
\frac{11840}{2402400}
\bigg\}.\label{cacfnfred}
\end{eqnarray}
With the same basis as in previous case~(\ref{Basiscf2nf}) it is easy to find the following general expression for Eq.~(\ref{cacfnfred}) for even values of $\M$
\begin{equation}\label{cacfnfredMP}
\widetilde{\gamma}_{\mathrm{TR},\mathrm{ns},\ca\cf\nf}^{\,(2)+}(\M)=
-\frac{4}{3}\ca\cf\nf\left(
\frac{\HS_1}{\M(\M+1)}
-\frac{22}{3}\frac{1}{\M(\M+1)}
\right)
\end{equation}
and for odd values of $\M$
\begin{equation}\label{cacfnfredMM}
\widetilde{\gamma}_{\mathrm{TR},\mathrm{ns},\ca\cf\nf}^{\,(2)-}(\M)=
4\,\ca\cf\nf \left(
\frac{\HS_1}{\M(\M+1)}
+\frac{8}{9}\frac{1}{\M(\M+1)}
-\frac{4}{3}\frac{1}{(\M+1)^2}
\right).
\end{equation}
The final expression for $\ca\cf\nf$ contribution can be found from Eqs.~(\ref{gammahatcacfnf}), (\ref{cacfnfredMP}) and (\ref{cacfnfredMM}):
\begin{equation}\label{cf2nfz3TR}
{\gamma}_{\mathrm{TR,ns},\ca\cf\nf}^{\,(2)\pm}(\M)=
\widehat{\gamma}_{\mathrm{ns},\ca\cf\nf}^{\,(2)}(\M)+
\widetilde{\gamma}_{\mathrm{TR,ns},\ca\cf\nf}^{\,(2)\pm}(\M)\,.
\end{equation}

\subsection{$\cf^3$ contribution}\label{subsec:cf3}

For $\cf^3$ contribution we again start with substitution $\Nminus\to 1$ and $\Nplus\to 1$ into the result for the three-loop non-singlet anomalous dimension from Ref.~\cite{Moch:2004pa}  and obtain
\begin{eqnarray}
\widehat{\gamma}_{\mathrm{ns},\cf^3}^{\,(2)}(\M)
&=& 16\,\cf^3 \bigg(
-\frac{29}{32}
+\frac38 \HS_2
+\frac{13}{4} \HS_3
-4 \HS_5
+3 \HS_{-2}
+9 \HS_{-4}
+12 \HS_{-5}\nonumber\\[1mm]&&\hspace*{-10mm}
-24 \HS_{-4,1}
-4 \HS_{-3,-2}
+6 \HS_{-3,1}
-4 \HS_{-3,2}
-12 \HS_{-2,-3}
-44 \HS_{1,-4}
-6 \HS_{1,-3}\nonumber\\[1mm]&&\hspace*{-10mm}
-12 \HS_{1,3}
+8 \HS_{1,4}
-52 \HS_{2,-3}
-6 \HS_{2,-2}
-6 \HS_{2,2}
-4 \HS_{2,3}
-12 \HS_{3,-2}
-8 \HS_{3,2}\nonumber\\[1mm]&&\hspace*{-10mm}
-4 \HS_{4,1}
+24 \HS_{-2,1,-2}
+64 \HS_{1,-3,1}
-16 \HS_{1,-2,-2}
-12 \HS_{1,-2,1}
+8 \HS_{1,-2,2}\nonumber\\[1mm]&&\hspace*{-10mm}
+80 \HS_{1,1,-3}
+16 \HS_{1,2,-2}
+8 \HS_{1,2,2}
+16 \HS_{1,3,1}
+56 \HS_{2,-2,1}
+16 \HS_{2,1,-2}\nonumber\\[1mm]&&\hspace*{-10mm}
+8 \HS_{2,1,2}
+8 \HS_{2,2,1}
+8 \HS_{3,1,1}
-96 \HS_{1,1,-2,1}
-\bigg(
\frac{15}{2}
+12 \HS_{-2}
\bigg)\,\z3
\bigg)\,.\label{gammahatcf3}
\end{eqnarray}
Subtracting this equation at $\M=1,\ldots,15$ from calculated values Eqs.~(\ref{res1})-(\ref{res15}) we obtain
\begin{eqnarray}
&\cf^3&\bigg\{
\frac{244}{3}-136 \,\z3\,,
76-24 \,\z3\,,
-\frac{10}{3}-24 \,\z3\,,
\frac{1909}{72}-\frac{16}{3}\,\z3\,,
-\frac{137437}{27000}-\frac{48}{5}\,\z3\,,\nonumber\\[2mm]&&\ \
\frac{31939}{2250}-\frac{12}{5}\,\z3\,,
-\frac{113172503}{27783000}-\frac{36}{7}\,\z3\,,
\frac{1330189999}{148176000}-\frac{48}{35}\,\z3\,,
-\frac{843934429}{266716800}-\frac{16}{5}\,\z3\,,\nonumber\\[2mm]&&\ \
\frac{1950321041}{312558750}-\frac{8}{9}\,\z3\,,
-\frac{2081091038207}{832031392500}-\frac{24}{11}\,\z3\,,
\frac{2456339454271}{532500091200}-\frac{48}{77}\,\z3\,,\nonumber\\[2mm]&&\ \
-\frac{532264770130798741}{263228107582440000}-\frac{144}{91}\,\z3\,,
\frac{729203803284910783}{204732972564120000}-\frac{6}{13}\,\z3\,,\nonumber\\[2mm]&&\ \
-\frac{34163492121141653}{20473297256412000} - \frac{6}{5}\,\z3
\bigg\}
\label{cf3red}\,.
\end{eqnarray}

Let's first try to reconstruct $\z3$ part.
It is clear from the last equation, that the possible basis has transcendentality level {1}. However the basis from $\cf\n2f$ part~(\ref{Basisnf2}) does not work.
Then, we slightly extended this basis
\begin{equation}\label{Basiscf3z3}
\mathcal{B}_{\cf^3,\z3}=
\left\{
\HS_1\,,\
\frac{1}{\M-1}\,,\
\frac{1}{\M}\,,\
\frac{1}{\M+1}\,,\
\frac{1}{\M+2}
\right\}
\end{equation}
and we have found for even values of $\M$
\begin{equation}\label{cf3z3redMP}
\widetilde{\gamma}_{\mathrm{TR},\mathrm{ns},\cf^3\z3}^{\,(2)+}(\M)=
32\,\cf^3\z3\left(
\frac{1}{\M-1}
-\frac{1}{\M+2}
\right)=
\frac{96\,\cf^3\z3}{(\M-1)(\M+2)}
\ ,
\end{equation}
while for odd values of $\M$
\begin{equation}\label{cf3z3redMM}
\widetilde{\gamma}_{\mathrm{TR},\mathrm{ns},\cf^3\z3}^{\,(2)-}(\M)=
288\,\cf^3\z3\left(
\frac{1}{\M}
-\frac{1}{\M+1}
\right)=
\frac{288\,\cf^3\z3}{\M(\M+1)}
\ .
\end{equation}

For construction of the basis for the rational part we analyze again the denominators with \texttt{FactorInteger}, for example:
\begin{equation}
204732972564120000 =2^6\times 3^6\times5^4 \times7^4 \times11^3 \times13^3.
\end{equation}
So, we assume, that the basis for the reduced $\cf^3$ part (\ref{cf3red}) should contain the basis for $\z3\,\cf^3$ part (\ref{Basiscf3z3}), multiplied by harmonic sums $\HS_{\vec m}(N)$ up to weight {{3}}. Moreover, we assume, that these harmonic sums should satisfy a generalized Gribov-Lipatov reciprocity~\cite{Dokshitzer:2005bf,Dokshitzer:2006nm}, which means, that up to weight {{3}} only the following combinations of the harmonic sums, called reciprocity-respecting harmonic sums~\cite{Dokshitzer:2006nm,Beccaria:2009vt}~\footnote{Note, that the reciprocity-respecting harmonic sums are equivalent to the binomial sums~\cite{Vermaseren:1998uu,Lukowski:2009ce}.}, should be considered:
\begin{equation}\label{HS3RR}
\Big\{\HS_1,\ \HS_{-2},\ \HS_1^2,\ \HS_3,\ \HS_1^3,\ \HS_1\HS_{-2},\ 2\HS_{-2,1}\!-\HS_{-3}
\Big\}.
\end{equation}
Thus, we stop with the following basis containing 18 terms:
\begin{eqnarray}\label{Basiscf3}
\mathcal{B}_{\cf^3}&=&
\bigg\{
\frac{1}{\M (\M+1)}\,,
\frac{1}{(\M-1)^3 (\M+2)^3}\,,
\frac{1}{(\M-1)^2 (\M+2)^2}\,,
\frac{1}{(\M-1) (\M+2)}\,,\nonumber\\[2mm]&&
\frac{\HS_{-2}}{\M (\M+1)}\,,
\frac{\HS_{-2}}{(\M-1) (\M+2)}\,,
\frac{\HS_1}{\M(\M+1)}\,,
\frac{\HS_1}{(\M-1) (\M+2)}\,,
\frac{\HS_{-2} \HS_1}{\M (\M+1)}\,,\nonumber\\[2mm]&&
\frac{\HS_{-2} \HS_1}{(\M-1) (\M+2)}\,,
\frac{\HS_1^2}{\M (\M+1)}\,,
\frac{\HS_1^2}{(\M-1) (\M+2)}\,,
\frac{\HS_1^3}{\M(\M+1)}\,,
\frac{\HS_1^3}{(\M-1) (\M+2)}\,,\nonumber\\[2mm]&&
\frac{\HS_3}{\M (\M+1)}\,,
\frac{\HS_3}{(\M-1) (\M+2)}\,,
\frac{2 \HS_{-2,1}-\HS_{-3}}{\M (\M+1)}\,,
\frac{2 \HS_{-2,1}-\HS_{-3}}{(\M-1) (\M+2)}
\bigg\}.
\end{eqnarray}
However, we have only 7 even values of anomalous dimension (\ref{cf3red}), from which we exclude $N=2$ moment, because multigluon operator starts from $N=4$. To find 18 coefficients from 6 known values we have used the same method, which we applied for the reconstruction of a general form of six-loop anomalous dimension of twist-3 operators in the maximally extended $\cN=4$ SYM theory~\cite{Velizhanin:2010cm}. Major observation that the coefficients in any anomalous dimension are usually the rather simple numbers, that is the equation for the coefficients is a Diophantine equation. Moreover, a lot of harmonic sums, which can be written down for the possible basis, absent in final expression, i.e. its coefficients are zeros. There is a nice algorithm, which can help to solve such problem. \texttt{LLL}-algorithm~\cite{Lenstra:1982} is realized in many computer algebra systems and its usage for our purpose can be found with the help of \texttt{MATHEMATICA}, where such algorithm is realized  with function \texttt{LatticeReduce}.
Firstly, we calculate the values of all 18 terms in the basis~(\ref{Basiscf3}) up to $N=14$.
So, we have $6$ equations in the linear system for $18$ variables.
We eliminate $5$ variables and we remain with one equation on $13$ variables.
According to the realization of \texttt{LLL}-algorithm\footnote{See {\texttt{Application}} at
{\texttt{{http://reference.wolfram.com/mathematica/ref/LatticeReduce.html}}}} we add to the $14\times 14$ unity matrix the column with numbers from the remaining equation of our system ($13 + 1$ numbers) and \texttt{MATHEMATICA} gives:
\begin{eqnarray}
\widetilde{\gamma}_{\mathrm{TR,ns},\cf^3}^{\,(2)+}(N) &=&
16\cf^3\bigg(
\frac{6 \HS_1}{\M (\M+1)}
+\frac{6 \HS_1}{(\M-1) (\M+2)}
+\frac{\HS_{-2}}{\M (\M+1)}
-\frac{6 \HS_{-2}}{(\M-1) (\M+2)}\nonumber\\&& \qquad
-\frac{6 \HS_{-2} \HS_1}{\M (\M+1)}
+\frac{6 \HS_{-2} \HS_1}{(\M-1)(\M+2)}
-\frac{ 6 \left(2 \HS_{-2,1}-\HS_{-3}\right)}{(\M-1) (\M+2)}
\bigg)\,.\label{cf3redMP}
\end{eqnarray}
Note, that \texttt{LatticeReduce} gives $14$ possible solutions but all other solutions contain large integers and only few zeros.

For odd values we should add to the basis few terms, which can be generated by Gribov-Lipatov reciprocity through a general equation~\cite{Dokshitzer:2005bf,Dokshitzer:2006nm}:
\begin{equation}\label{RReq}
\gamma(N)={\mathcal P}\Big(N+\sigma\gamma(N)\Big)\,,
\end{equation}
where $\sigma$ is some numerical factor and ${\mathcal P}(N)$ translated into the $x$-space is the reciprocity respecting function~\cite{Dokshitzer:2005bf,Dokshitzer:2006nm}. This equation means, that the part of anomalous dimension in $n$-loop order is generated by the low order results.
A new term proportional to $1/N/(N+1)$ appears at the first time namely for odd values of anomalous dimension at two-loop order (see the last line of Eq.~(\ref{eq:gqqTR1p})). This term will generate at third order the contribution, which looks like
\begin{equation}\label{RRgen}
\frac{d}{dN}\left(\frac{\HS_{1}(N)}{N(N+1)}\right).
\end{equation}
So, we should add to the basis~(\ref{Basiscf3}) the following terms with harmonic sums:
\begin{equation}\label{BasisOdd}
\bigg\{
\frac{\HS_{2}}{N(N+1)}\,,
\frac{\HS_{1}}{N^2}\,,
\frac{\HS_{1}}{(N+1)^2}
\bigg\}\ .
\end{equation}
Indeed, for this extended basis \texttt{LLL}-algorithm gives the following general expression for odd numbers from Eq.~(\ref{cf3red}):
\begin{eqnarray}
\widetilde{\gamma}_{\mathrm{TR},\mathrm{ns},\cf^3}^{\,(2)-}(N) &=&
16\cf^3\bigg(
\frac{4}{N (N+1)}
+\frac{1}{(N-1) (N+2)}
+\frac{3}{(N+1)^2}
+\frac{2 \HS_1}{N^2 (N+1)^2}\nonumber\\&&\hspace*{-15mm}
-\frac{4 \HS_1}{(N+1)^2}
+\frac{5 \HS_1}{N (N+1)}
-\frac{2 \HS_1}{(N-1)(N+2)}
+\frac{2 \HS_2}{N (N+1)}
+\frac{\HS_{-2}}{(N-1) (N+2)}\nonumber\\&&\hspace*{-15mm}
-\frac{2 \HS_{-2} \HS_1}{(N-1) (N+2)}
+\frac{18 \HS_{-2} \HS_1}{N (N+1)}
-\frac{18 (2\HS_{-2,1}-\HS_{-3})}{N (N+1)}
\bigg)\,.\label{cf3redMM}
\end{eqnarray}

The final expression for $\cf^3$ contribution can be found from Eqs.~(\ref{gammahatcf3}), (\ref{cf3z3redMP}), (\ref{cf3z3redMM}), (\ref{cf3redMP}) and (\ref{cf3redMM}) as
\begin{equation}\label{cf3TR}
{\gamma}_{\mathrm{TR,ns},\cf^3}^{\,(2)\pm}(N)=
\widehat{\gamma}_{\mathrm{ns},\cf^3}^{\,(2)}(N)+
\widetilde{\gamma}_{\mathrm{TR,ns},\cf^3}^{\,(2)\pm}(N)+
\widetilde{\gamma}_{\mathrm{TR,ns},\cf^3\z3}^{\,(2)\pm}(N)\,.
\end{equation}

\subsection{$\ca\cf^2$ contribution}\label{subsec:cacf2}

Substitute $\Nminus\to 1$ and $\Nplus\to 1$ into the result for non-singlet anomalous dimension from Ref.~\cite{Moch:2004pa} we obtain for $\ca\cf^2$ contribution
\begin{eqnarray}
\widehat{\gamma}_{\mathrm{ns},\ca\cf^2}^{\,(2)}(N)&=& 16\ca\cf^2 \bigg(
-\frac{151}{64}
+ \frac{151}{24} \HS_2
- \frac{13}{2} \HS_3
+ \frac{23}{6} \HS_4
+ 10 \HS_5
- \frac{9}{2} \HS_{-2}\nonumber\\&&\hspace*{-15mm}
+ \frac{134}{9} \HS_{-3}
- \frac{89}{6} \HS_{-4}
- 10 \HS_{-5}
+ 20 \HS_{-4,1}
-2 \HS_{-3,-2}
-\frac{31}{3} \HS_{-3,1}
+2 \HS_{-3,2}\nonumber\\&&\hspace*{-15mm}
+10 \HS_{-2,-3}
-6 \HS_{-2,-2}
+34 \HS_{1,-4}
+\frac{31 }{3}\HS_{1,-3}
-\frac{268 }{9}\HS_{1,-2}
-\frac{134 }{9}\HS_{1,2}\nonumber\\&&\hspace*{-15mm}
+\frac{62 }{3}\HS_{1,3}
-22 \HS_{1,4}
+42 \HS_{2,-3}
+\frac{31}{3}\HS_{2,-2}
-\frac{134}{9}\HS_{2,1}
+\frac{22}{3}\HS_{2,2}
-10 \HS_{2,3}\nonumber\\&&\hspace*{-15mm}
+10 \HS_{3,-2}
+\frac{4}{3}\HS_{3,1}
-2 \HS_{3,2}
-2 \HS_{4,1}
+8 \HS_{-2,-2,1}
-28 \HS_{-2,1,-2}
-64 \HS_{1,-3,1}\nonumber\\&&\hspace*{-15mm}
+32 \HS_{1,-2,-2}
+\frac{62}{3} \HS_{1,-2,1}
-4 \HS_{1,-2,2}
-72 \HS_{1,1,-3}
+16 \HS_{1,1,3}
-8 \HS_{1,2,-2}\nonumber\\&&\hspace*{-15mm}
-16 \HS_{1,3,1}
-60 \HS_{2,-2,1}
-8 \HS_{2,1,-2}
+112 \HS_{1,1,-2,1}
+\frac{3}{2} \left(
\frac{15}{2}
+12 \HS_{-2}
\right)\z3
\bigg)\label{cf2caV}\,.
\end{eqnarray}
Subtracting this equation at $N=1,\ldots,15$ from calculated values Eqs.~(\ref{res1})-(\ref{res15}) we obtain
\begin{eqnarray}
&\ca\cf^2&\bigg\{
-\frac{1424}{9}+220 \z3\,,
-\frac{283}{3}+36 \z3\,,
\frac{5}{18}+36 \z3\,,
-\frac{4477}{144}+8 \z3\,,
\frac{94739}{18000}+\frac{72\z3}{5}\,,\nonumber\\[3mm]&&
-\frac{36817}{2250}+\frac{18 \z3}{5}\,,
\frac{28787477}{6174000}+\frac{54 \z3}{7}\,,
-\frac{3040082519}{296352000}+\frac{72 \z3}{35}\,,
\frac{2007772853}{533433600}+\frac{24 \z3}{5}\,,\nonumber\\[4mm]&&
-\frac{11821223651}{1666980000}+\frac{4 \z3}{3}\,,
\frac{20222794198253}{6656251140000}+\frac{36 \z3}{11}\,,
-\frac{27807644469347}{5325000912000}+\frac{72 \z3}{77}\,,\nonumber\\[4mm]&&
\frac{1310584038208950541}{526456215164880000}+\frac{216 \z3}{91}\,,
-\frac{1646199071310197233}{409465945128240000}+\frac{9 \z3}{13}\,,\nonumber\\[4mm]&&
\frac{8492728741110641}{4094659451282400} + \frac{9 \z3}{5}
\bigg\}
\label{cf2cared}\,.
\end{eqnarray}
Using the same basis (\ref{Basiscf3}) as in previous subsection we have found with \texttt{LLL}-algorithm the following general expression for even numbers from Eq.~(\ref{cf2cared})
\begin{eqnarray}
\widetilde{\gamma}_{\mathrm{TR},\mathrm{ns},\ca\cf^2}^{\,(2)+}(N) &=&
8\,\ca\cf^2\bigg(
-\frac{103 \HS_1}{6 N (N+1)}
-\frac{18 \HS_1}{(N-1) (N+2)}
-\frac{3 \HS_{-2}}{N (N+1)}
\nonumber\\[2mm]&&\hspace*{-20mm}
+\frac{18 \HS_{-2}}{(N-1) (N+2)}
-\frac{18 \HS_{-2} \HS_1}{(N-1) (N+2)}
+\frac{6 \HS_{-2} \HS_1}{N (N+1)}
+\frac{18 (2\HS_{-2,1}-\HS_{-3})}{(N-1) (N+2)}
\bigg)\,.\qquad\label{cacf2redMP}
\end{eqnarray}
For odd numbers from Eq.~(\ref{cf2cared}) we extend basis (\ref{Basiscf3}) with few additional terms~(\ref{BasisOdd}) and obtain
\begin{eqnarray}
\widetilde{\gamma}_{\mathrm{TR},\mathrm{ns},\ca\cf^2}^{\,(2)-}(N) &=&
8\,\ca\cf^2\bigg(
-\frac{31}{3 (N+1)^2}
-\frac{209}{9 N (N+1)}
-\frac{3}{(N-1) (N+2)}
+\frac{4 \HS_1}{(N+1)^2}
\nonumber\\&&\hspace*{-20mm}
-\frac{2 \HS_1}{N^2 (N+1)^2}
-\frac{53 \HS_1}{6 N (N+1)}
+\frac{2 \HS_1}{(N-1)(N+2)}
-\frac{2 \HS_2}{N (N+1)}
\nonumber\\&&\hspace*{-20mm}
-\frac{3 \HS_{-2}}{(N-1) (N+2)}
+\frac{2 \HS_{-2} \HS_1}{(N-1) (N+2)}
-\frac{54 \HS_{-2} \HS_1}{N (N+1)}
+\frac{54 (2\HS_{-2,1}-\HS_{-3})}{N (N+1)}
\bigg)\,.\label{cacf2redMM}
\end{eqnarray}

For $\z3$ part the results are the same as in $\cf^3$ case in Eqs.~(\ref{cf3z3redMP}) and (\ref{cf3z3redMM}) up to common factor~$(-3/2)$.
The final expression for $\ca\cf^2$ contribution can be found from Eqs.~(\ref{cf2caV}), (\ref{cacf2redMP}), (\ref{cacf2redMM}), (\ref{cf3z3redMP}) and (\ref{cf3z3redMM}) as
\begin{equation}\label{cacf2TR}
{\gamma}_{\mathrm{TR,ns},\ca\cf^2}^{\,(2)\pm}(N)=
 \widehat{\gamma}_{\mathrm{ns},\ca\cf^2}^{\,(2)}(N)
+\widetilde{\gamma}_{\mathrm{TR,ns},\ca\cf^2}^{\,(2)\pm}(N)
-\frac32\,\frac{\ca\cf^2}{\cf^3}\, \widetilde{\gamma}_{\mathrm{TR,ns},\cf^3\z3}^{\,(2)\pm}(N)\ .
\end{equation}

\subsection{$\ca^2\cf$ contribution}\label{subsec:ca2cf}

For the last $\ca^2\cf$ contribution our reduction rule $\Nminus\to 1$ and $\Nplus\to 1$ for the result of the three-loop non-singlet anomalous dimension from Ref.~\cite{Moch:2004pa} gives
\begin{eqnarray}
\widehat{\gamma}_{\mathrm{ns},\ca^2\cf}^{\,(2)+}(N)&=&
16\,\ca^2 \cf \bigg(
\frac{1657}{576}
+\frac{245 }{24}\HS_1
-\frac{1043}{108} \HS_2
+\frac{389}{36} \HS_3
-\frac{31}{6} \HS_4
-2 \HS_5\nonumber\\&&\hspace*{-15mm}
+\frac32 \HS_{-2}
-\frac{67}{9} \HS_{-3}
+\frac{31}{6} \HS_{-4}
+2 \HS_{-5}
-4 \HS_{-4,1}
+2 \HS_{-3,-2}
+\frac{11}{3} \HS_{-3,1}\nonumber\\&&\hspace*{-15mm}
-2 \HS_{-2,-3}
+3 \HS_{-2,-2}
-6 \HS_{1,-4}
-\frac{11}{3} \HS_{1,-3}
+\frac{134}{9} \HS_{1,-2}
-\frac{11}{2} \HS_{1,3}
+6 \HS_{1,4}\nonumber\\&&\hspace*{-15mm}
-8 \HS_{2,-3}
-\frac{11}{3}\HS_{2,-2}
+4 \HS_{2,3}
-2 \HS_{3,-2}
+\frac{11}{6} \HS_{3,1}
-2 \HS_{4,1}
-4 \HS_{-2,-2,1} \nonumber\\&&\hspace*{-15mm}
+8 \HS_{-2,1,-2}
+16 \HS_{1,-3,1}
-12 \HS_{1,-2,-2}
-\frac{22}{3} \HS_{1,-2,1}
+16 \HS_{1,1,-3}
-8 \HS_{1,1,3}\nonumber\\&&\hspace*{-15mm}
+8 \HS_{1,3,1}
+16 \HS_{2,-2,1}
-32 \HS_{1,1,-2,1}
-\frac{1}{2}\,
\bigg(
\frac{15}{2}
+12 \HS_{-2}
\bigg)\,\z3
\bigg)\label{ca2cfV}\,.
\end{eqnarray}
Subtracting this equation at $N=1,\ldots,15$ from calculated values Eqs.~(\ref{res1})-(\ref{res15}) we obtain
\begin{eqnarray}
&&\ca^2\cf
\bigg\{
\frac{125}{3}-76 \z3\,,
\frac{590}{27}-12 \z3\,,
-\frac{8}{3}-12 \z3\,,
\frac{817}{120}-\frac{8 \z3}{3}\,,
-\frac{15193}{5400}-\frac{24 \z3}{5}\,,\nonumber\\[2mm]&&\qquad\quad
\frac{134749}{37800}-\frac{6\z3}{5}\,,
-\frac{846463}{396900}-\frac{18 \z3}{7}\,,
\frac{14190767}{6350400}-\frac{24 \z3}{35}\,,
-\frac{10288711}{6350400}-\frac{8 \z3}{5}\,,\nonumber\\[2mm]&&\qquad\quad
\frac{1803287}{1166400}-\frac{4 \z3}{9}\,,
-\frac{972218609}{768398400}-\frac{12\z3}{11}\,,
\frac{85376995477}{74918844000}-\frac{24 \z3}{77}\,,\nonumber\\[2mm]&&\qquad\quad
-\frac{593181857171}{584366983200}-\frac{72 \z3}{91}\,,
\frac{4789546276259}{5454091843200}-\frac{3\z3}{13}\,,
-\frac{7574874351263}{9090153072000} - \frac{3 \z3}{5}
\bigg\}.\qquad\quad
\label{cfca2red}
\end{eqnarray}
Using the same basis (\ref{Basiscf3}) as in subsection~\ref{subsec:cf3} we have found with \texttt{LLL}-algorithm the following general expression for even numbers from Eq.~(\ref{cfca2red})
\begin{eqnarray}
\widetilde{\gamma}_{\mathrm{TR},\mathrm{ns},\ca^2\cf}^{\,(2)+}(N) &=&
\ca^2\cf
\bigg(
-\frac{121}{9 N (N+1)}
+\frac{56 \HS_1}{3 N (N+1)}
+\frac{24 \HS_1}{(N-1) (N+2)}
\nonumber\\&&\hspace*{-20mm}
+\frac{4 \HS_{-2}}{N (N+1)}
-\frac{24 \HS_{-2}}{(N-1) (N+2)}
+\frac{24 \HS_{-2} \HS_1}{(N-1) (N+2)}
-\frac{24 (2\HS_{-2,1}-\HS_{-3})}{(N-1) (N+2)}
\bigg)\,,\qquad\label{ca2cfredMP}
\end{eqnarray}
while for odd numbers from Eq.~(\ref{cfca2red}) we obtain with extended basis from Eqs.~(\ref{Basiscf3}) and~(\ref{BasisOdd})
\begin{eqnarray}
\widetilde{\gamma}_{\mathrm{TR},\mathrm{ns},\ca^2\cf}^{\,(2)-}(N) &=&
\ca^2\cf
\bigg(
-\frac{44}{3 (N+1)^2}
-\frac{25}{N (N+1)}
-\frac{4}{(N-1) (N+2)}
\nonumber\\&&\hspace*{-15mm}
-\frac{4 \HS_1}{N (N+1)}
-\frac{4 \HS_{-2}}{(N-1) (N+2)}
-\frac{72 \HS_{-2} \HS_1}{N (N+1)}
+\frac{72 (2\HS_{-2,1}-\HS_{-3})}{N (N+1)}
\bigg)\,.\quad\label{ca2cfredMM}
\end{eqnarray}

For $\z3$ part the results are the same as for $\cf^3$ contribution in Eqs.~(\ref{cf3z3redMP}) and (\ref{cf3z3redMM}) up to common factor~$1/2$.
The final expression for $\ca^2\cf$ contribution can be found from Eqs.~(\ref{ca2cfV}), (\ref{ca2cfredMP}), (\ref{ca2cfredMM}), (\ref{cf3z3redMP}) and (\ref{cf3z3redMM}) as
\begin{equation}\label{ca2cfTR}
{\gamma}_{\mathrm{TR,ns},\ca^2\cf}^{\,(2)\pm}(N)=
  \widehat{\gamma}_{\mathrm{ns},\ca^2\cf}^{\,(2)}(N)
+ \widetilde{\gamma}_{\mathrm{TR,ns},\ca^2\cf}^{\,(2)\pm}(N)
+ \frac{1}{2}\,\frac{\ca\cf^2}{\cf^3}\, \widetilde{\gamma}_{\mathrm{TR,ns},\cf^3\z3}^{\,(2)\pm}(N)\ .
\end{equation}

\setcounter{equation}{0}
\section{Results in Mellin space}\label{sec:ResultsN}
Here we present the final results for the three-loop anomalous dimensions of the flavour non-singlet transversity operator~(\ref{op2}).
For representation of our results we use the same notation as in Ref.~\cite{Moch:2004pa}
\begin{equation}
\label{eq:shiftN}
  \Npm \, S_{\vec{m}}  =  S_{\vec{m}}(N \pm 1)\,, \quad\quad
  \Npmi\, S_{\vec{m}}  =  S_{\vec{m}}(N \pm i)
\end{equation}
for arguments shifted by $\pm 1$ or a larger integer $i$.
In this notation the one-loop (LO) anomalous dimension~\cite{Baldracchini:1980uq,Shifman:1980dk,Bukhvostov:1985rn,Artru:1989zv,Blumlein:2001ca} reads
\begin{eqnarray}
  \gamma^{\,(0)}_{\,\mathrm {TR,ns}}(N) & = &
         {\cf} \* \big(
            4 \* \S(1)
          - 3
          \big)\label{eq:gqq0}
\end{eqnarray}
and the corresponding second-order (NLO) non-singlet quantities~\cite{Hayashigaki:1997dn,Kumano:1997qp,Vogelsang:1997ak} are given by
\begin{eqnarray}
  &&\hspace*{-2mm}\gamma^{\,(1)+}_{\,\mathrm {TR,ns}}(N) =
  4\, \* {\ca \* \cf} \* \bigg(
          - {17 \over 24}
          + {67 \over 9} \* \S(1)
          - {11 \over 3} \* \S(2)
          + 2 \* \S(3)
          - 2 \* \S(-3)
          + 4 \* \Ss(1,-2)
          \bigg)
  \nonumber\\&& \
+ 4\, \* {\cf \* \nf} \* \bigg(
            {1 \over 12}
          - {10 \over 9} \* \S(1)
          + {2 \over 3} \* \S(2)
          \bigg)
+ 4\, \* {\cf^{2}} \* \bigg(
           3 \* \S(2)
          - {3 \over 8}
          + 4 \* \S(-3)
          - 8 \* \Ss(1,-2)
          - 4 \* \Ss(1,2)
          - 4 \* \Ss(2,1)
          \bigg)\!,\qquad \label{eqgqq1p}
 \\[2mm]
  &&\hspace*{-2mm}\gamma^{\,(1)-}_{\,\mathrm {TR,ns}}(N)  =
     \gamma^{\,(1)+}_{\,\mathrm {TR,ns}}(N)
       + 8\, \*  {\cf \* \bigg(\cf - {\ca \over 2} \bigg)}\,
           (\Nminus+\Nplus-2) \* \S(1)\ .
 \label{eqgqq1m}
\end{eqnarray}

The three-loop (NNLO, N$^{\,2}$LO) contribution to the anomalous dimension $\gamma^{\,+}_{\,\mathrm {TR,ns}}(N)$ for even moments reads from Eqs.~(\ref{gammahatcfnf2}), (\ref{gammahatcf2nf}), (\ref{gammahatcacfnf}), (\ref{gammahatcf3}), (\ref{cf2caV}), (\ref{ca2cfV})
as
\begin{eqnarray}
&&{\gamma}_{\mathrm{TR,ns}}^{\,(2)+}(N)
=
\widehat{\gamma}_{\mathrm{ns},\cf\n2f}^{\,(2)+}(N)
+\widehat{\gamma}_{\mathrm{ns},\cf^2\nf}^{\,(2)+}(N)
+\widehat{\gamma}_{\mathrm{ns},\ca\cf\nf}^{\,(2)+}(N)
+\widehat{\gamma}_{\mathrm{ns},\cf^3}^{\,(2)+}(N)
\nonumber\\&&\quad
+\widehat{\gamma}_{\mathrm{ns},\ca\cf^2}^{\,(2)+}(N)
+\widehat{\gamma}_{\mathrm{ns},\ca^2\cf}^{\,(2)+}(N)
+ \frac{8}{9}\*\cf\*\nf^2\*(\Nminus + \Nplus - 2)\*\HS_{1}
\nonumber\\&&\quad
+ 16\cf^2\*\nf\*\bigg(
 \frac{1}{3}\*(\Nplus-\N0)\*\HS_{2}
-\frac{1}{3}\*(\Nminus + \Nplus-2)\*\HS_{1, 1}
\bigg)
\nonumber\\&&\quad
+16\ca\*\cf\*\nf\*\bigg(
- \frac{1}{12}\*(\Nplus-\N0)\*\HS_{2}
+ (\Nminus + \Nplus - 2)\*\bigg[
- \frac{11}{18}\*\HS_{1}
+ \frac{1}{12}\*\HS_{1, 1}\bigg]
\bigg)
\nonumber\\&&\quad
+32\*\cf\*(\cf-\ca)\*\bigg(\cf - \frac{\ca}{2}\bigg)
\*(\Nminustwo -\Nminus- \Nplus + \Nplustwo)\*\HS_{1} \*\z3\
\nonumber\\&&\quad
+16\ca^2\*\cf\*\bigg(
 (\Nplus-\N0)\*\bigg[
  \frac{10}{3}\*\HS_{2}
  - \frac{1}{2}\*\HS_{3}
  \bigg]
+(\Nminus + \Nplus-2)\*\bigg[
  \frac{121}{72}\*\HS_{1}
  - \frac{10}{3}\*\HS_{1, 1}
  + \frac{1}{2}\*\HS_{1, -2}
  \bigg]
\nonumber\\&&\qquad
-(\Nminus - \Nplus)\*\HS_{2, 1}
+(\Nplustwo-\Nplus)\*\bigg[
  \HS_{2, -2}
  - \HS_{3, 1}
  \bigg]
\nonumber\\&&\qquad
+(\Nminustwo - \Nminus - \Nplus + \Nplustwo)\*\bigg[
  \HS_{1, -2, 1}
  - \HS_{1, 1, -2}
  \bigg]
  \bigg)
\nonumber\\&&\quad
+ 16\ca\*\cf^2\*\bigg(
3\*(\Nminus - \Nplus)\*\HS_{2, 1}
+ (\Nplus-\N0)\*\bigg[
 - \frac{139}{12}\*\HS_{2}
 + \frac{3}{2}\*\HS_{3}
 + 3\*\HS_{4}
 + 3\*\HS_{2, -2}
 - 3\*\HS_{3, 1}
\bigg]
\nonumber\\&&\qquad
+ 3(\Nplustwo-\Nplus)\*\bigg[
    \*\HS_{3, 1}
  - \*\HS_{2, -2}
  \bigg]
+ (\Nminus + \Nplus-2)\*\bigg[
    \frac{139}{12}\*\HS_{1, 1}
  - \frac{3}{2}\*\HS_{1, -2}
  + 3\*\HS_{1, -3}
\nonumber\\&&\qquad
  - 3\*\HS_{1, -2, 1}
  - 3\*\HS_{1, 1, -2}
  \bigg]
+ 3(\Nminustwo-\Nminus - \Nplus + \Nplustwo)\*\bigg[
   \*\HS_{1, 1, -2}
 - \*\HS_{1, -2, 1}
 \bigg]
\bigg)
\nonumber\\&&\quad
+ 16\cf^3\*\bigg(
- 2\*(\Nminus - \Nplus)\*\HS_{2, 1}
+ 2\*(\Nplustwo-\Nplus)\*\bigg[\HS_{2, -2} - \HS_{3, 1}\bigg]
\nonumber\\&&\qquad
+ (\Nplus-\N0)\*\bigg[8\*\HS_{2} - \HS_{3} - 6\*\HS_{4} - 6\*\HS_{2, -2} + 6\*\HS_{3, 1}\bigg]
\nonumber\\&&\qquad
+ (\Nminus + \Nplus-2)\*\bigg[
  - 8\*\HS_{1, 1}
  + \HS_{1, -2}
  - 6\*\HS_{1, -3}
  + 6\*\HS_{1, -2, 1}
  + 6\*\HS_{1, 1, -2}
  \bigg]
\nonumber\\&&\qquad
+ 2\*(\Nminustwo-\Nminus  - \Nplus + \Nplustwo)\*\bigg[
   \HS_{1, -2, 1}
   - \HS_{1, 1, -2}
   \bigg]
\bigg).\label{eqgqq2p}
\end{eqnarray}
\begin{figure}[t]
  \includegraphics[width=8cm]{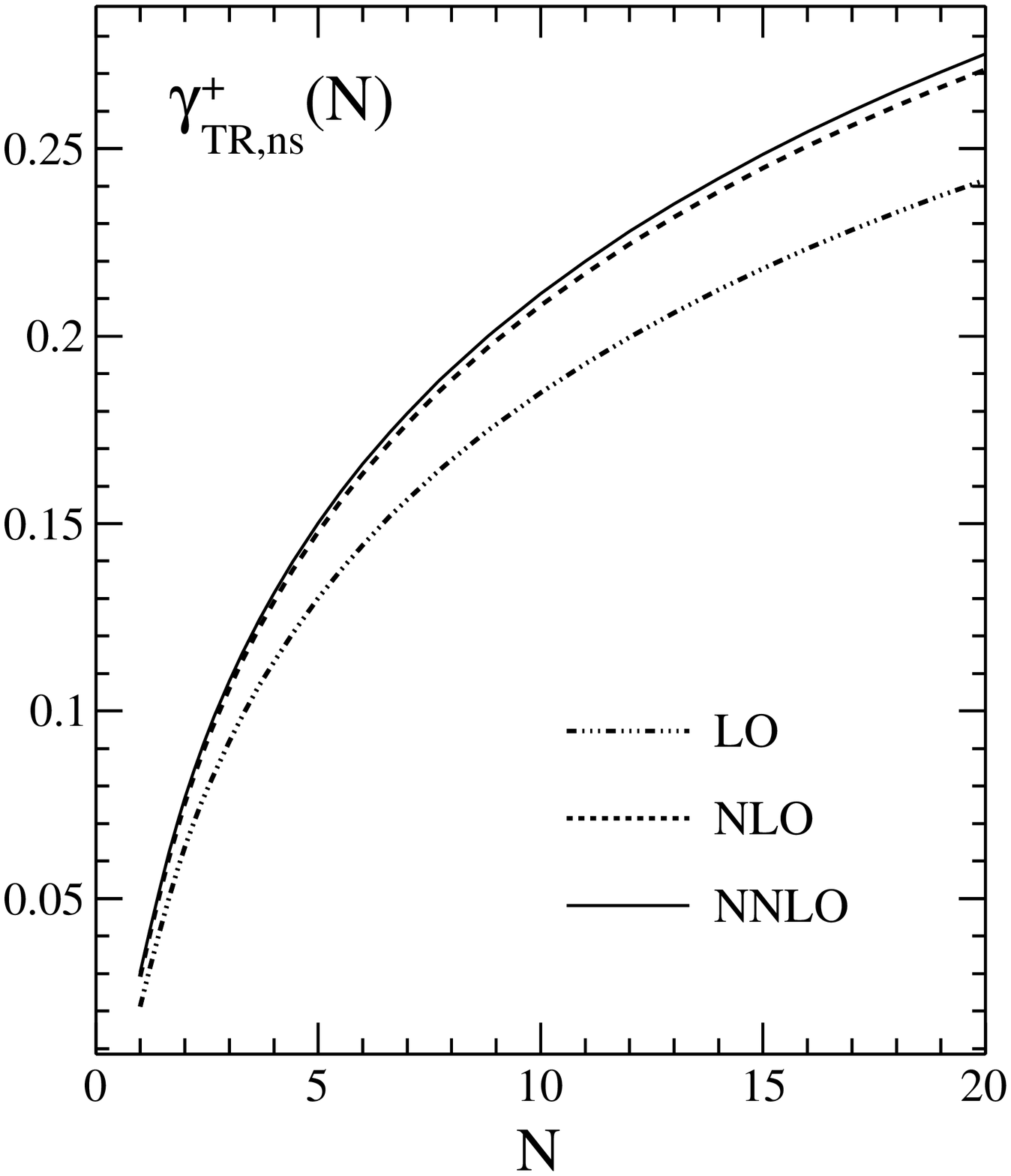}
  \includegraphics[width=8cm]{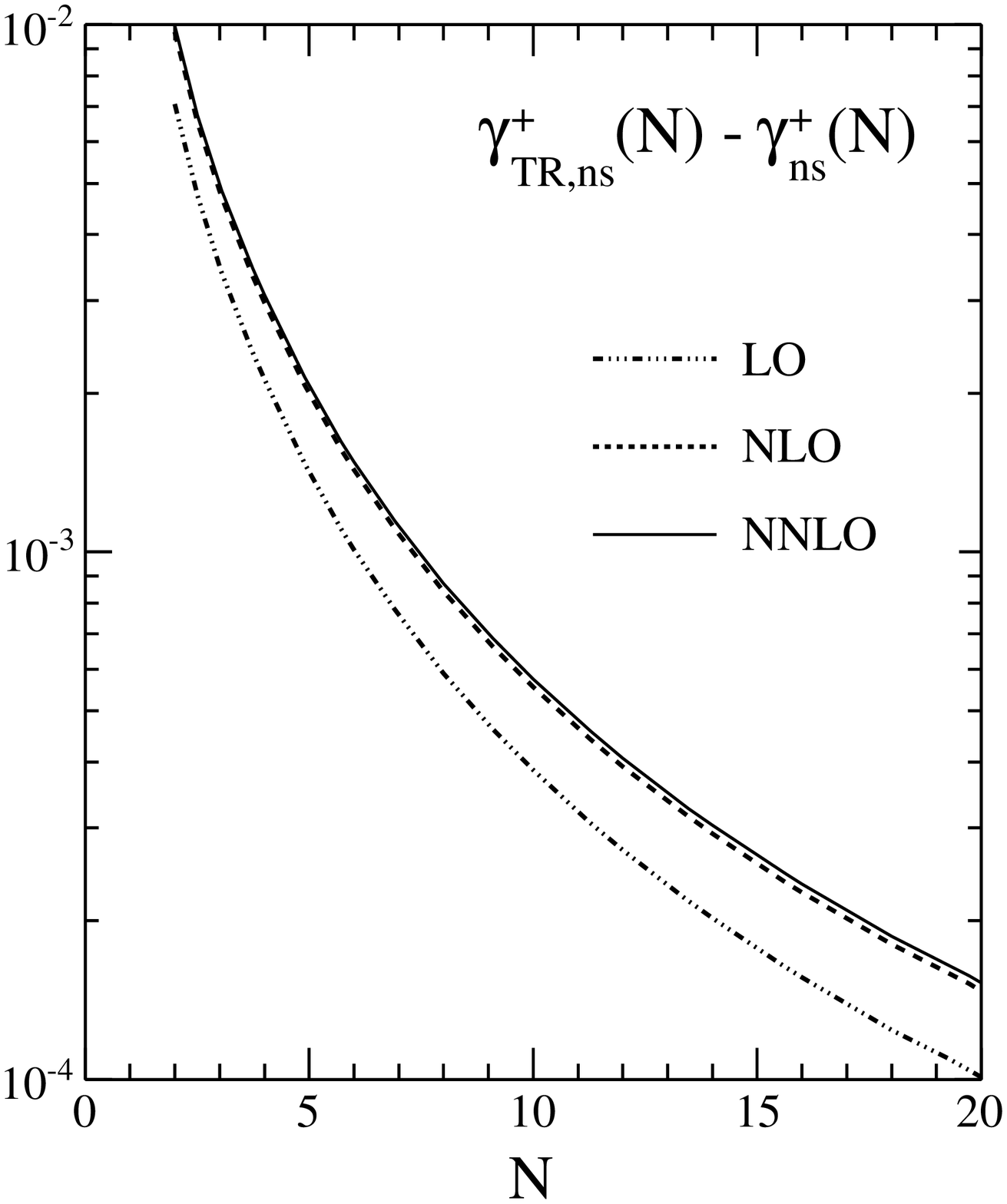}
\caption{The perturbative expansion of the anomalous dimension
 $\gamma_{\,{\mathrm {TR,ns}}}^{\, +}(N)$ for four flavours at $\alpha_s = 0.2$.
In the right part the difference between our three-loop result and the corresponding result from Ref.~\cite{Moch:2004pa} is shown.}
\label{fig:gplus}
\end{figure}
For odd moments we have
\begin{eqnarray}
 &&{\gamma}_{\mathrm{TR,ns}}^{\,(2)-}(\M)=
 {\gamma}_{\mathrm{TR,ns}}^{\,(2)+}(\M)
+ 16\cf\*\nf\*\bigg(\cf - \frac{\ca}{2}\bigg)\*(\Nminus + \Nplus - 2)\*
  \bigg[-\frac{7}{9}\*\HS_{1} + \frac{2}{3}\*\HS_{1, 1}\bigg] +
\nonumber\\&&\quad
+32\*\cf\*(\cf-\ca)\!\bigg(\cf - \frac{\ca}{2}\bigg)\!\Big(
    9\*(\Nminus + \Nplus - 2)\*\HS_{1} -
   (\Nminustwo - \Nminus - \Nplus + \Nplustwo)\*\HS_{1}
   \Big)\*\z3
\nonumber\\&&\quad
+ 16\ca\*\cf\*\bigg(\cf - \frac{\ca}{2}\bigg)\!\bigg(
   (\Nplus-\N0)\*\bigg(2\*\HS_{2} - \HS_{3} - 18\*\HS_{2, -2} + 18\*\HS_{3, 1}\bigg)
\nonumber\\&&\qquad
 + (\Nminus - \Nplus)\*\bigg[\frac{1}{3}\*\HS_{2} - 2\*\HS_{2, 1}\bigg]
 + (\Nplustwo-\Nplus)\*\bigg[\frac{1}{3}\*\HS_{3} + 2\*\HS_{2, -2} - 2\*\HS_{3, 1}\bigg]
\nonumber\\&&\qquad
 + (\Nminus + \Nplus - 2)\*\bigg[
       \frac{179}{18}\*\HS_{1}
       + \HS_{1, -2}
       - \frac{17}{3}\*\HS_{1, 1}
       - 18\*\HS_{1, -2, 1}
       + 18\*\HS_{1, 1, -2}
       \bigg]
\nonumber\\&&\qquad
 + (\Nminustwo - \Nminus - \Nplus + \Nplustwo)\*\bigg[
      \frac{1}{3}\*\HS_{1, -2}
    + 2\*\HS_{1, -2, 1}
    - 2\*\HS_{1, 1, -2}
    \bigg]
 \bigg)
\nonumber\\&&\quad
+ 16\cf^2\*\bigg(\cf - \frac{\ca}{2}\bigg)\*
\bigg(
(\Nplus-\N0)\*\bigg[-4\*\HS_{2} + \frac{19}{3}\*\HS_{3} + 6\*\HS_{4} +
     24\*\HS_{2, -2} - 4\*\HS_{2, 1} - 24\*\HS_{3, 1}\bigg]
\nonumber\\&&\qquad
+\frac{2}{3}\*(\Nminus - \Nplus)\*\HS_{2, 1}
+(\Nplustwo-\Nplus)\*\bigg[\frac{1}{3}\*\HS_{3} - \frac{2}{3}\*\HS_{4} -
     \frac{8}{3}\*\HS_{2, -2} + \frac{8}{3}\*\HS_{3, 1}\bigg]
\nonumber\\&&\qquad
+(\Nminustwo - \Nminus - \Nplus + \Nplustwo)\*\bigg[
   - \frac{2}{3}\*\HS_{1, -3}
   + \frac{1}{3}\*\HS_{1, -2}
   - \frac{4}{3}\*\HS_{1, -2, 1}
   + \frac{8}{3}\*\HS_{1, 1, -2}
   \bigg]
\nonumber\\&&\qquad
+(\Nminus + \Nplus-2)\*\bigg[
   - \frac{13}{3}\*\HS_{1}
   - \frac{1}{3}\*\HS_{2}
   - \frac{2}{3}\*\HS_{3}
   + 6\*\HS_{1, -3}
   - \frac{5}{3}\*\HS_{1, -2}
   + \frac{23}{3}\*\HS_{1, 1}
\nonumber\\&&\qquad
   - 2\*\HS_{1, 2}
   + 12\*\HS_{1, -2, 1}
   - 24\*\HS_{1, 1, -2}\bigg]
\bigg).\label{eqgqq2m}
\end{eqnarray}

The results (\ref{eq:gqq0}), (\ref{eqgqq1p}) and (\ref{eqgqq2p}) for $\gamma_{\mathrm {TR,ns}}^{\, +}(N)$ are collected in Fig.~\ref{fig:gplus} for four active
flavours and a typical value $\alpha_s = 0.2$ for the strong coupling constant as in Ref.~\cite{Moch:2004pa}. Numerically, the colour factors take
the values $\cf=4/3, \ca=3$. Moreover, in the right part we show the difference between $\gamma_{\mathrm {TR,ns}}^{\, +}(N)$ and $\gamma_{\mathrm {ns}}^{\, +}(N)$ from Ref.~\cite{Moch:2004pa} and one can see, that the difference between NLO and NNLO results is very small.

\setcounter{equation}{0}
\section{Results in x-space}\label{sec:ResultsX}

The non-singlet anomalous dimensions $\gamma_{\,\mathrm {TR,ns}}^{\,(n)\pm}(\M)$ are related by the Mellin transformation
\begin{equation}
\label{eq:Pdef}
  \gamma_{\,\mathrm {TR,ns}}^{(n)\pm}(\M)  =
  - \int_0^1 \!dx\, x^{\,\M-1}\, P_{\,\mathrm {TR,ns}}^{\,(n)\pm}(x)
\end{equation}
to the splitting functions $P_{\,\mathrm {TR,ns}}^{\,(n)\pm}(x)$.
So, the splitting functions $P_{\,\mathrm {TR,ns}}^{\,(n)\pm}(x)$ can be obtained from the $N$-space results of the previous section by an inverse Mellin
transformation, which expresses these functions in terms of harmonic polylogarithms~\cite{Goncharov,Borwein:1999js,Remiddi:1999ew}.
The inverse Mellin transformation can be performed by a completely algebraic procedure with HARMPOL package~\cite{Vermaseren:1998uu,Remiddi:1999ew} for
FORM~\cite{Vermaseren:2000nd}.

Our notation for the harmonic polylogarithms $H_{m_1,...,m_w}(x)$, $m_j = 0,\pm 1$ follows Ref.~\cite{Remiddi:1999ew} to which the reader is referred for a detailed discussion.
The lowest-weight ($w = 1$) functions $H_m(x)$ are given by
\begin{equation}
\label{eq:hpol1}
  H_0(x)        =  \ln x \, , \quad\quad
  H_{\pm 1}(x)  =  \mp \, \ln (1 \mp x) \, .
\end{equation}
The higher-weight ($w \geq 2$) functions are recursively defined as
\begin{equation}
\label{eq:hpol2}
  H_{m_1,...,m_w}(x)  =
    \left\{ \begin{array}{cl}
    \displaystyle{ \frac{1}{w!}\,\ln^w x \, ,}
       & \quad {\rm if} \, m^{}_1,...,m^{}_w = 0,\ldots ,0 \\[2ex]
    \displaystyle{ \int_0^x \! dz\, f_{m_1}(z) \, H_{m_2,...,m_w}(z)
       \,, } & \quad {\rm else}
    \end{array} \right.
\end{equation}
with
\begin{equation}
\label{eq:hpolf}
  f_0(x)       =  \frac{1}{x} \, , \quad\quad
  f_{\pm 1}(x)  =  \frac{1}{1 \mp x} \, .
\end{equation}
A useful short-hand notation is
\vspace*{-3mm}
\begin{equation}
\label{eq:habbr}
  H_{{\scriptsize \underbrace{0,\ldots ,0}_{\scriptstyle m} },\,
  \pm 1,\, {\scriptsize \underbrace{0,\ldots ,0}_{\scriptstyle n} },
  \, \pm 1,\, \ldots}(x)  =  H_{\pm (m+1),\,\pm (n+1),\, \ldots}(x)
  \, .
\end{equation}
\vspace*{-3mm}

The one-loop non-singlet splitting function~\cite{Baldracchini:1980uq,Shifman:1980dk,Bukhvostov:1985rn,Artru:1989zv,Blumlein:2001ca} can be written as
\begin{equation}
  P^{\,(0)}_{\,\mathrm {TR,ns}}(x) =
  {\cf} \* \Big( \,
          2 \* \pqq(x)
          + 3\* \delta(1 - x)
          \,\Big),
\label{eq:Pqq0}
\end{equation}
where
\begin{equation}
  \pqqx  = \frac{2\,x}{1 - x}\ .
\end{equation}
Here and below we suppress the argument $x$ of the polylogarithms and all divergences for $x \to 1 $ are understood in the sense of $+$-distributions.

The two-loop non-singlet splitting functions~\cite{Hayashigaki:1997dn,Kumano:1997qp,Vogelsang:1997ak} are given by
\begin{eqnarray}
&&P^{\,(1)+}_{\,\mathrm {TR,ns}}(x) =
  4\cf\*\nf\*\pqqx\*\bigg[
      - \frac{5}{9}
      - \frac{1}{3}\*\HP_{0}
      +  \delta(1-x)\* \bigg(
            {1 \over 12}
          + {2 \over 3} \* \z2
          \bigg)
\bigg]
  \nonumber\\&&\quad
+ 4\ca\*\cf\*\bigg(
           \pqqx\*\bigg[
                   \frac{67}{18}
                 - \z2
                 + \frac{11}{6}\*\HP_{0}
                 + \HP_{0, 0}
                 \bigg]
         + \pqqmx\*\bigg[
                   \z2
                 + 2\*\HP_{-1, 0}
                 - \HP_{0, 0}
                 \bigg]
          \bigg)
  \nonumber\\&&\quad
+ 4\cf^2\*\bigg(
          2\pqqx\*\bigg[
                   \*\HP_{1, 0}
                 - \frac{3}{4}\*\HP_{0}
                 + \*\HP_{2}
                \bigg]
        - 2\pqqmx\*\bigg[
                   \z2
                 + 2\*\HP_{-1, 0}
                 - \HP_{0, 0}
                 \bigg]
        \bigg)
  \nonumber\\&&\quad
       + 4\delta(1 - x) \* \bigg[
         \ca\cf\bigg(
            {17 \over 24}
          + {11 \over 3} \* \z2
          - 3 \* \z3
          \bigg)
       + \cf^2 \* \bigg(
           {3 \over 8}
          - 3 \* \z2
          + 6 \* \z3
          \bigg)
          \bigg]
,\\[3mm]
&&P^{\,(1)-}_{\,\mathrm {TR,ns}}(x) =P^{\,(1)+}_{\,\mathrm {TR,ns}}(x)
  \nonumber\\&&\quad
+16\cf\left(\cf-\frac{\ca}{2}\right)\*\bigg(
         \pqqmx\*\bigg[
                      \z2
                    + 2\*\HP_{-1, 0}
                    - \HP_{0, 0}
                    \bigg]
           - \frac{1}{2}\*\Big(1-x\Big)
           \bigg).
\end{eqnarray}

The three-loop splitting function for the evolution corresponding to the anomalous dimension $\gamma^{(2)+}_{\,\mathrm {TR,ns}}(\M)$ from Eq.~(\ref{eqgqq2p}) reads
\begin{eqnarray}
&&P^{\,(2)+}_{\,\mathrm {TR,ns}}(x) =
16\*\cf\*\nf^2\*
\bigg(
\frac{1}{18}\*(1 - x) + \frac{1}{27}
+ \pqqx\*\bigg[
- \frac{1}{54\,x}
+ \frac{5}{54}\*\HP_{0}
+ \frac{1}{18}\*\HP_{0, 0}
\bigg]
\bigg)
\nonumber\\&&\quad
+16\*\ca\*\cf\*\nf\*
\bigg(
- \frac{x}{12}\*\HP_{0}
+ (1 - x)\*\bigg[
     - \frac{11}{18}
     + \frac{1}{12}\*\HP_{1}
\bigg]
+ \pqqmx\*\bigg[
   \frac{1}{2}\*\z3
 - \frac{5}{9}\*\z2
 - \frac{2}{3}\*\z2\*\HP_{-1}
\nonumber\\&&\quad
 + \frac{1}{6}\*\z2\*\HP_{0}
 - \frac{1}{3}\*\HP_{3}
 - \frac{1}{3}\*\HP_{-2, 0}
 - \frac{10}{9}\*\HP_{-1, 0}
 + \frac{2}{3}\*\HP_{-1, 2}
 + \frac{5}{9}\*\HP_{0, 0}
 - \frac{1}{3}\*\HP_{-1, 0, 0}
 + \frac{1}{3}\*\HP_{0, 0, 0}
 \bigg]
\nonumber\\&&\quad
+ \pqqx\*\bigg[
   \frac{5}{9}\*\z2
 - \frac{3}{2}\*\z3
 - \frac{209}{216}
 + \left(\frac{1}{3}\*\z2 - \frac{167}{108}\right)\*\HP_{0}
 - \frac{1}{6}\*\HP_{3}
 - \frac{7}{6}\*\HP_{0, 0}
 - \frac{1}{3}\*\HP_{0, 0, 0}
 + \frac{1}{2}\*\HP_{1, 0, 0}
 \bigg]
\bigg)
\nonumber\\&&\quad
+16\*\cf^2\*\nf\*
\bigg(
   \frac{x}{3}\*\HP_{0}
 - \frac{1}{3}\*(1 - x)\*\HP_{1}
 + \pqqmx\*\bigg[
     \frac{10}{9}\*\z2
   - \z3
   + \frac{4}{3}\*\z2\*\HP_{-1}
   - \frac{1}{3}\*\z2\*\HP_{0}
   + \frac{2}{3}\*\HP_{3}
\nonumber\\&&\quad
   + \frac{2}{3}\*\HP_{-2, 0}
   + \frac{20}{9}\*\HP_{-1, 0}
   - \frac{4}{3}\*\HP_{-1, 2}
   - \frac{10}{9}\*\HP_{0, 0}
   + \frac{2}{3}\*\HP_{-1, 0, 0}
   - \frac{2}{3}\*\HP_{0, 0, 0}
   \bigg]
\nonumber\\&&\quad
 + \pqqx\*\bigg[
     \frac{5}{3} \*\z3
   - \frac{55}{48}
   + \left(\frac{1}{3}\*\z2 + \frac{5}{24}\right)\*\HP_{0}
   - \frac{10}{9}\*\HP_{2}
   - \frac{2}{3}\*\HP_{3}
   + \frac{1}{2}\*\HP_{0, 0}
   - \frac{10}{9}\*\HP_{1, 0}
   - \frac{2}{3}\*\HP_{2, 0}
\nonumber\\&&\quad
   - \frac{1}{3}\*\HP_{0, 0, 0}
   - \frac{4}{3}\*\HP_{1, 0, 0}
   \bigg]
\bigg)
+16\*\cf^3\*
\bigg(
  \HP_{0, 0}
- 3\*\z3
- 8\*\HP_{0}
+ 6\*\z2\*\HP_{0}
- 6\*\HP_{3}
- 6\*\HP_{-2, 0}
+ 6\*\HP_{0, 0, 0}
\nonumber\\&&\quad
+ x^2\*\Big[-\z3 - 2\*\z2\*\HP_{0} - 2\*\z2\*\HP_{1} + 2\*\HP_{3} + 2\*\HP_{-2, 0}\Big]
+(1 - x)\*\bigg[-6\*\z3 - \frac{1}{2}\*\z2 - 9\*\z2\*\HP_{0} - 8\*\HP_{1}
\nonumber\\&&\quad
- 3\*\z2\*\HP_{1} + 6\*\HP_{3} - \HP_{0, 0}\bigg]
+ \left(\frac{1}{x} + x^2\right)\*\bigg[\z2\*\HP_{-1} + \z2\*\HP_{1} - 2\*\HP_{-1, 2} -   2\*\HP_{-1, -1, 0}\bigg]
\nonumber\\&&\quad
+ (1 + x)\*\bigg[9\*\z3 - \frac{3}{2}\*\z2 - 9\*\z2\*\HP_{-1} +
   8\*\HP_{0} + 3\*\z2\*\HP_{0} + 2\*\HP_{2} + 6\*\HP_{-2, 0} + \HP_{-1, 0} +
   6\*\HP_{-1, 2}
\nonumber\\&&\quad
   - 6\*\HP_{-1, -1, 0} + 6\*\HP_{-1, 0, 0} - 6\*\HP_{0, 0, 0}\bigg]
+ \pqqmx\*\bigg[
    \frac{7}{2}\*\z2^2
  - \frac{9}{2}\*\z3
  + 32\*\z2\*\HP_{-2}
  + 6\*\z2\*\HP_{2}
  + 3\*\HP_{3}
\nonumber\\&&\quad
  + \big(6\*\z2 + 36\*\z3\big)\*\HP_{-1}
  + \left(-13\*\z3 - \frac{3}{2} - \frac{3}{2}\*\z2\right)\*\HP_{0}
  + 12\*\HP_{4}
  - 6\*\HP_{-3, 0}
  + 3\*\HP_{-2, 0}
  - 28\*\HP_{-2, 2}
\nonumber\\&&\quad
  - 48\*\z2\*\HP_{-1, -1}
  + 40\*\z2\*\HP_{-1, 0}
  - 6\*\HP_{-1, 2}
  - 32\*\HP_{-1, 3}
  - 14\*\z2\*\HP_{0, 0}
  + 2\*\HP_{3, 0}
  + 8\*\HP_{-2, -1, 0}
\nonumber\\&&\quad
  - 26\*\HP_{-2, 0, 0}
  + 8\*\HP_{-1, -2, 0}
  + 48\*\HP_{-1, -1, 2}
  + 3\*\HP_{-1, 0, 0}
  - 4\*\HP_{-1, 2, 0}
  - \frac{9}{2}\*\HP_{0, 0, 0}
  + 40\*\HP_{-1, -1, 0, 0}
\nonumber\\&&\quad
  - 22\*\HP_{-1, 0, 0, 0}
  + 6\*\HP_{0, 0, 0, 0}
  \bigg]
+ \pqqx\*\bigg[
    \frac{9}{10}\*\z2^2
  + 6\*\z2\*\HP_{-2}
  + \left(\z3 - \frac{3}{2}\*\z2 - \frac{3}{16}\right)\*\HP_{0}
  + 12\*\z3\*\HP_{1}
\nonumber\\&&\quad
  + 2\*\HP_{4}
  - 2\*\HP_{-3, 0}
  + \frac{13}{8}\*\HP_{0, 0}
  + 8\*\HP_{1, 3}
  - 3\*\HP_{2, 0}
  + 4\*\HP_{2, 2}
  + 4\*\HP_{3, 0}
  + 4\*\HP_{3, 1}
  + 12\*\HP_{-2, -1, 0}
\nonumber\\&&\quad
  - 6\*\HP_{-2, 0, 0}
  + 8\*\HP_{1, -2, 0}
  - 6\*\HP_{1, 0, 0}
  + 4\*\HP_{1, 2, 0}
  + 2\*\HP_{2, 0, 0}
  + 4\*\HP_{2, 1, 0}
  - 2\*\HP_{0, 0, 0, 0}
  - 4\*\HP_{1, 0, 0, 0}
  \bigg]
\bigg)
\nonumber\\&&\quad
+16\ca^2\*\cf\*
\bigg(
-\frac{10}{3}\*\HP_{0}
+ \frac{1}{2}\*\HP_{0, 0}
+ x^2\*\Big[-\frac{1}{2}\*\z3 - \z2\*\HP_{0} - \z2\*\HP_{1} + \HP_{3} + \HP_{-2, 0}\Big]
\nonumber\\&&\quad
+ (1 + x)\*\bigg[-\frac{3}{4}\*\z2 + \frac{10}{3}\*\HP_{0} + \HP_{2} +
   \frac{1}{2}\*\HP_{-1, 0}\bigg]
+ (1 - x)\*\bigg[\frac{121}{72} - \frac{1}{4}\*\z2 -
   \frac{10}{3}\*\HP_{1} - \frac{1}{2}\*\HP_{0, 0}\bigg]
\nonumber\\&&\quad
+ \left(\frac{1}{x} + x^2\right)\*\bigg[\frac{1}{2}\*\z2\*\HP_{-1} + \frac{1}{2}\*\z2\*\HP_{1} -
   \HP_{-1, 2} - \HP_{-1, -1, 0}\bigg]
+ \pqqmx\*\bigg[
    \frac{67}{18}\*\z2
  - \z2^2
  - \frac{11}{4}\*\z3
\nonumber\\&&\quad
  + 8\*\z2\*\HP_{-2}
  + \left(12\*\z3 + \frac{11}{3}\*\z2\right)\*\HP_{-1}
  + \left(-4\*\z3 - \frac{1}{6}\*\z2 - \frac{3}{4}\right)\*\HP_{0}
  + 2\*\z2\*\HP_{2}
  + \frac{11}{6}\*\HP_{3}
  + 2\*\HP_{4}
\nonumber\\&&\quad
  - \HP_{-3, 0}
  + \frac{11}{6}\*\HP_{-2, 0}
  - 8\*\HP_{-2, 2}
  - 16\*\z2\*\HP_{-1, -1}
  + \left(11\*\z2 + \frac{67}{9}\right)\*\HP_{-1, 0}
  - \frac{11}{3}\*\HP_{-1, 2}
  - 8\*\HP_{-1, 3}
\nonumber\\&&\quad
  + \left(-3\*\z2 - \frac{67}{18}\right)\*\HP_{0, 0}
  - 4\*\HP_{-2, 0, 0}
  + 16\*\HP_{-1, -1, 2}
  + \frac{11}{6}\*\HP_{-1, 0, 0}
  - \frac{31}{12}\*\HP_{0, 0, 0}
  + 8\*\HP_{-1, -1, 0, 0}
\nonumber\\&&\quad
  - 3\*\HP_{-1, 0, 0, 0}
  + \HP_{0, 0, 0, 0}\bigg]
+ \pqqx\*\bigg[
    \frac{1}{2}\*\z3
  + \frac{12}{5}\*\z2^2
  - \frac{67}{18}\*\z2
  + \frac{245}{48}
  + 9\*\z3\*\HP_{1}
  + \frac{11}{12}\*\HP_{3}
  + \HP_{4}
\nonumber\\&&\quad
  + \left(4\*\z3 - \frac{31}{12}\*\z2 + \frac{1043}{216}\right)\*\HP_{0}
  + \HP_{-3, 0}
  - \frac{3}{2}\*\HP_{-2, 0}
  + 2\*\HP_{-2, 2}
  + \frac{389}{72}\*\HP_{0, 0}
  - \z2\*\HP_{1, 0}
  + 4\*\HP_{1, 3}
\nonumber\\&&\quad
  + 4\*\HP_{-2, -1, 0}
  - \HP_{-2, 0, 0}
  + \frac{31}{12}\*\HP_{0, 0, 0}
  + 6\*\HP_{1, -2, 0}
  - \frac{11}{4}\*\HP_{1, 0, 0}
  - 2\*\HP_{2, 0, 0}
  - \HP_{0, 0, 0, 0}
  - 3\*\HP_{1, 0, 0, 0}
\nonumber\\&&\quad
  - 4\*\HP_{1, 1, 0, 0}
  \bigg]
\bigg)
+16\ca\*\cf^2\*
\bigg(
  \frac{3}{2}\*\z3
- 3\*\z2\*\HP_{0}
+ \frac{139}{12}\*\HP_{0}
+ 3\*\HP_{3}
+ 3\*\HP_{-2, 0}
- \frac{3}{2}\*\HP_{0, 0}
- 3\*\HP_{0, 0, 0}
\nonumber\\&&\quad
+ x^2\*\bigg[\frac{3}{2}\*\z3 + 3\*\z2\*\HP_{0} + 3\*\z2\*\HP_{1} - 3\*\HP_{3} -
   3\*\HP_{-2, 0}\bigg]
+ (1 - x)\*\bigg[3\*\z3 + \frac{3}{4}\*\z2 + \frac{9}{2}\*\z2\*\HP_{0} +
   \frac{3}{2}\*\z2\*\HP_{1}
\nonumber\\&&\quad
   + \frac{139}{12}\*\HP_{1} - 3\*\HP_{3} +
   \frac{3}{2}\*\HP_{0, 0}\bigg]
+ \left(\frac{1}{x} + x^2\right)\*\bigg[-\frac{3}{2}\*\z2\*\HP_{-1}
       - \frac{3}{2}\*\z2\*\HP_{1} + 3\*\HP_{-1, 2} + 3\*\HP_{-1, -1, 0}\bigg]
\nonumber\\&&\quad
+ (1 + x)\*\bigg[
     \frac{9}{4}\*\z2
   - \frac{9}{2}\*\z3
   + \frac{9}{2}\*\z2\*\HP_{-1} - \frac{3}{2}\*\z2\*\HP_{0} -
   \frac{139}{12}\*\HP_{0} - 3\*\HP_{2} - 3\*\HP_{-2, 0} -
   \frac{3}{2}\*\HP_{-1, 0}
\nonumber\\&&\quad
   - 3\*\HP_{-1, 2}
   + 3\*\HP_{-1, -1, 0} -
   3\*\HP_{-1, 0, 0} + 3\*\HP_{0, 0, 0}\bigg]
+ \pqqmx\*\bigg[
     \frac{1}{4}\*\z2^2
   + \frac{31}{4}\*\z3
   - \frac{67}{9}\*\z2
\nonumber\\&&\quad
   - 32\*\z2\*\HP_{-2}
   + \left(-42\*\z3 - \frac{31}{3}\*\z2\right)\*\HP_{-1}
   + \left(\frac{9}{4} + \frac{13}{12}\*\z2 + \frac{29}{2}\*\z3\right)\*\HP_{0}
   - 7\*\z2\*\HP_{2}
   - \frac{31}{6}\*\HP_{3}
\nonumber\\&&\quad
   - 10\*\HP_{4}
   + 5\*\HP_{-3, 0}
   - \frac{31}{6}\*\HP_{-2, 0}
   + 30\*\HP_{-2, 2}
   + 56\*\z2\*\HP_{-1, -1}
   + \left(-42\*\z2 - \frac{134}{9}\right)\*\HP_{-1, 0}
\nonumber\\&&\quad
   + \frac{31}{3}\*\HP_{-1, 2}
   + 32\*\HP_{-1, 3}
   + \left(13\*\z2 + \frac{67}{9}\right)\*\HP_{0, 0}
   - \HP_{3, 0}
   - 4\*\HP_{-2, -1, 0}
   + 21\*\HP_{-2, 0, 0}
\nonumber\\&&\quad
   - 4\*\HP_{-1, -2, 0}
   - 56\*\HP_{-1, -1, 2}
   - \frac{31}{6}\*\HP_{-1, 0, 0}
   + 2\*\HP_{-1, 2, 0}
   + \frac{89}{12}\*\HP_{0, 0, 0}
   - 36\*\HP_{-1, -1, 0, 0}
\nonumber\\&&\quad
   + 17\*\HP_{-1, 0, 0, 0}
   - 5\*\HP_{0, 0, 0, 0}
   \bigg]
+ \pqqx\*\bigg[
     \frac{5}{6} \*\z3
   - \frac{69}{20} \*\z2^2
   + \left(-\frac{17}{2}\*\z3 + \frac{41}{12}\*\z2 - \frac{151}{48}\right)\*\HP_{0}
\nonumber\\&&\quad
   - 3\*\z2\*\HP_{-2}
   - 24\*\z3\*\HP_{1}
   + \left(-2\*\z2 + \frac{67}{9}\right)\*\HP_{2}
   + \frac{2}{3}\*\HP_{3}
   + \HP_{4}
   + \left(-4\*\z2 - \frac{13}{4}\right)\*\HP_{0, 0}
\nonumber\\&&\quad
   - \HP_{-3, 0}
   + 3\*\HP_{-2, 0}
   - 4\*\HP_{-2, 2}
   + \left(-2\*\z2 + \frac{67}{9}\right)\*\HP_{1, 0}
   - 8\*\HP_{1, 3}
   + \frac{11}{3}\*\HP_{2, 0}
   + \HP_{3, 0}
\nonumber\\&&\quad
   - 14\*\HP_{-2, -1, 0}
   + 5\*\HP_{-2, 0, 0}
   - \frac{23}{12}\*\HP_{0, 0, 0}
   - 16\*\HP_{1, -2, 0}
   + \frac{31}{3}\*\HP_{1, 0, 0}
   + 5\*\HP_{2, 0, 0}
   + 5\*\HP_{0, 0, 0, 0}
\nonumber\\&&\quad
   + 11\*\HP_{1, 0, 0, 0}
   + 8\*\HP_{1, 1, 0, 0}
   \bigg]
\bigg)
+ 16\,\delta(1 - x)\*\bigg[
 \cf\n2f \*  \bigg(
          - {17 \over 144}
          + {5 \over 27} \* \z2
          - {1 \over 9} \* \z3
          \bigg)
\nonumber\\[2mm]&&\quad
+ \cf^2\nf \*  \bigg(
            {5 \over 12} \* \z2
          - {23 \over 16}
          + {29 \over 30} \* \z2^2
          - {17 \over 6} \* \z3
          \bigg)
+ \ca\cf\nf \* \bigg(
            {5 \over 4}
          - {167 \over 54} \* \z2
          + {1 \over 20} \* \z2^2
          + {25 \over 18} \* \z3
          \bigg)
\nonumber\\[2mm]&&\quad
+ \ca\cf^2 \* \bigg(
            {151 \over 64}
          + \z2 \* \z3
          - {205 \over 24} \* \z2
          - {247 \over 60} \* \z2^2
          + {211 \over 12} \* \z3
          + {15 \over 2} \* \z5
          \bigg)
+ \ca^2\cf \*  \bigg(
            {281 \over 27} \* \z2
          -  {1657 \over 576}
\nonumber\\[2mm]&&\quad
          - {1 \over 8} \* \z2^2
          - {97 \over 9} \* \z3
          + {5 \over 2} \* \z5
          \bigg)
+ \cf^3\*  \bigg(
            {29 \over 32}
          - 2 \* \z2 \* \z3
          + {9 \over 8} \* \z2
          + {18 \over 5} \* \z2^2
          + {17 \over 4} \* \z3
          - 15 \* \z5
          \bigg)
\bigg]
\,.\label{eq:Pqq2p}
\end{eqnarray}
\begin{figure}[t]
  \includegraphics[width=8cm]{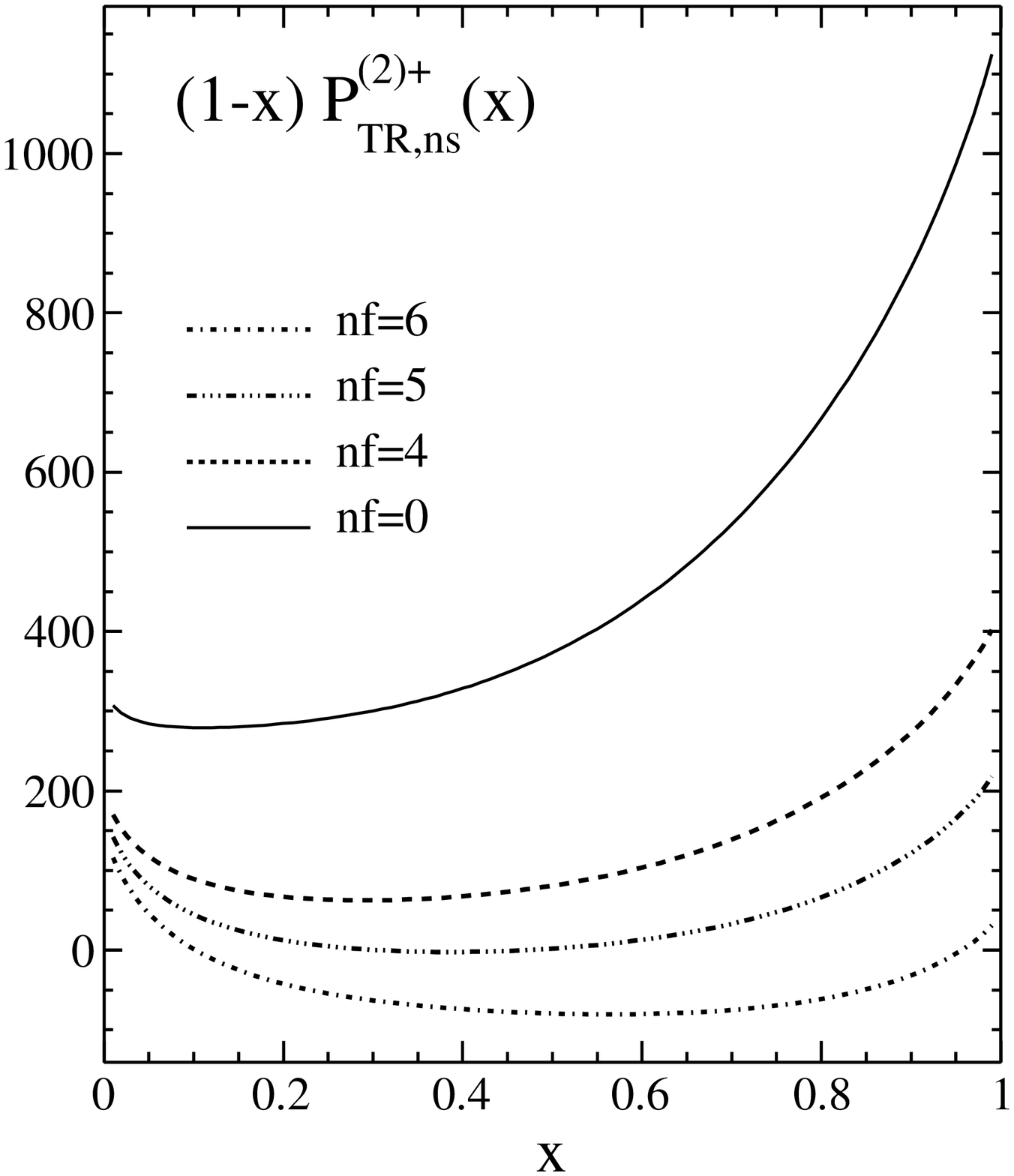}
  \includegraphics[width=8cm]{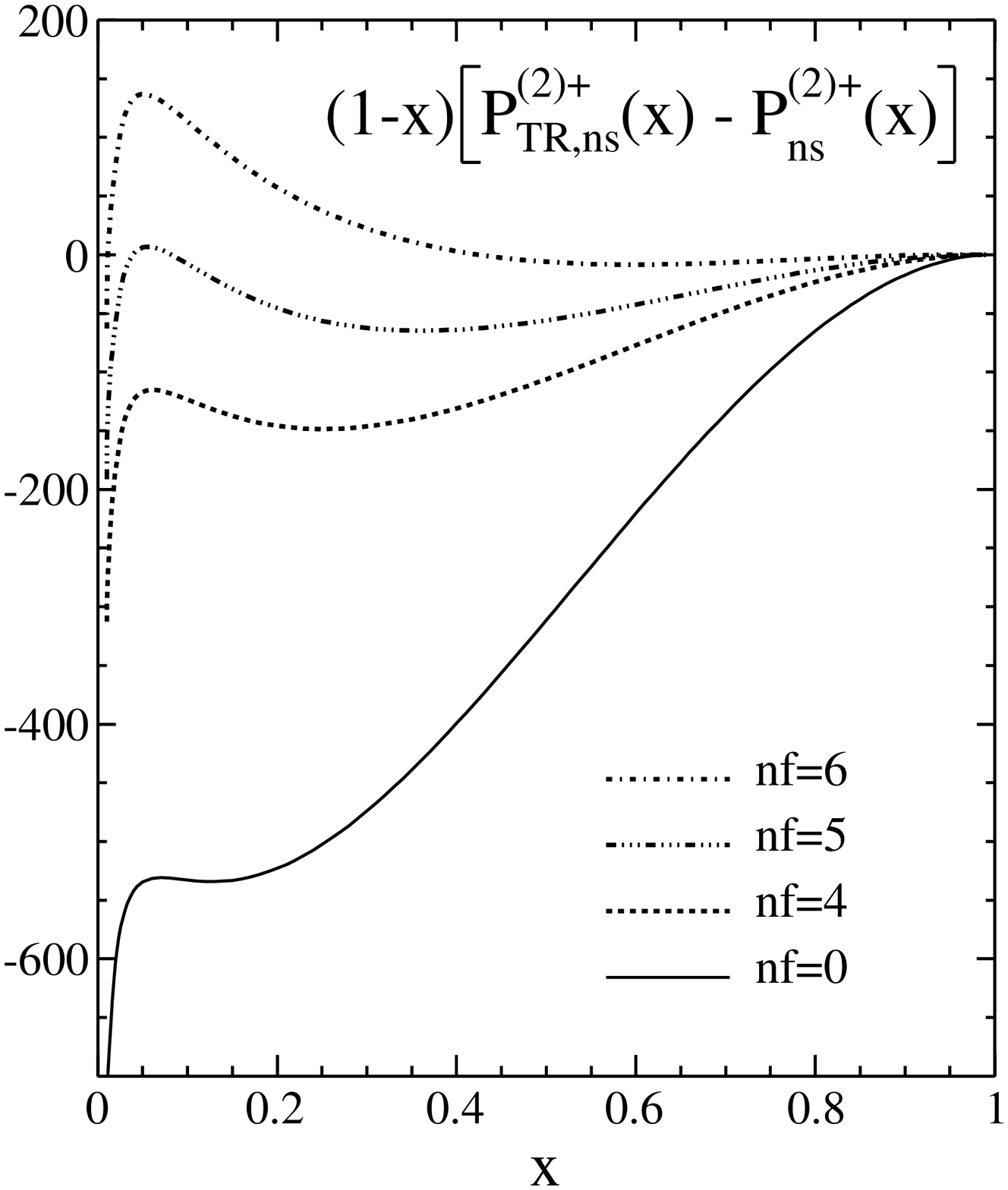}
\caption{The three-loop splitting function $P_{\mathrm {TR,ns}}^{\,(2) +}(x)$ for different numbers of active quarks,  multiplied by $(1-x)$ as in Ref.~\cite{Moch:2004pa}. In the right part the difference between our three-loop result and the corresponding result from Ref.~\cite{Moch:2004pa} is shown.}
\label{fig:PTRns2p}
\end{figure}
The function $P^{(2)+}_{{\mathrm {TR,ns}}}(x)$ from Eq.~(\ref{eq:Pqq2p}) is shown in Fig.~\ref{fig:PTRns2p} together with
the difference between our result and the result for $P^{(2)+}_{{\mathrm {ns}}}(x)$ from Ref.~\cite{Moch:2004pa}.
The numerical evaluations of the harmonic polylogarithm, entering into expression for $P^{(2)+}_{{\mathrm {TR,ns}}}(x)$ from Eq.~(\ref{eq:Pqq2p}), were performed with \texttt{HPL} package~\cite{Maitre:2005uu,Maitre:2007kp} for \texttt{MATHEMATICA}.

The three-loop splitting function for the evolution corresponding to the anomalous dimension $\gamma^{(2)-}_{\,\mathrm {TR,ns}}(N)$ from Eq.~(\ref{eqgqq2m}) is given by
\begin{eqnarray}
&&P^{\,(2)-}_{\,\mathrm {TR,ns}}(x) = P^{\,(2)+}_{\,\mathrm {TR,ns}}(x)
+16\*\cf\*\nf\*\left(\cf-\frac{\ca}{2}\right)\*
\bigg(
(1 - x)\*\bigg[-\frac{7}{9} + \frac{2}{3}\*\HP_{1}\bigg]
\nonumber\\&&\quad
+ \pqqmx\*\bigg[
     2\*\z3
   - \frac{20}{9}\*\z2
   - \frac{8}{3}\*\z2\*\HP_{-1}
   + \frac{2}{3}\*\z2\*\HP_{0}
   - \frac{4}{3}\*\HP_{3}
   - \frac{4}{3}\*\HP_{-2, 0}
   - \frac{40}{9}\*\HP_{-1, 0}
   + \frac{8}{3}\*\HP_{-1, 2}
\nonumber\\&&\quad
   + \frac{20}{9}\*\HP_{0, 0}
   - \frac{4}{3}\*\HP_{-1, 0, 0}
   + \frac{4}{3}\*\HP_{0, 0, 0}
   \bigg]
\bigg)
+16\*\cf^2\*\left(\cf-\frac{\ca}{2}\right)\*
\bigg(
- 4\*\z2
- 42\*\z3
- 42\*\z2\*\HP_{0}
\nonumber\\&&\quad
+ \frac{14}{3}\*\HP_{0}
+ 4\*\HP_{2}
+ 24\*\HP_{3}
- 12\*\HP_{-2, 0}
- \frac{19}{3}\*\HP_{0, 0}
- 6\*\HP_{0, 0, 0}
+ (1 - x)\*\bigg[33\*\z3 + \frac{5}{6}\*\z2 - \frac{13}{3}
\nonumber\\&&\quad
  + 36\*\z2\*\HP_{0}
  + 12\*\z2\*\HP_{1} + \frac{23}{3}\*\HP_{1} - 24\*\HP_{3} +
   \frac{17}{3}\*\HP_{0, 0} + 2\*\HP_{1, 0}\bigg]
+ \left(\frac{1}{x} + x^2\right)\*\bigg[-2\*\z2\*\HP_{-1}
\nonumber\\&&\quad
  - \frac{4}{3}\*\z2\*\HP_{1}
  + \frac{1}{3}\*\HP_{-1, 0} + \frac{8}{3}\*\HP_{-1, 2} +
   \frac{4}{3}\*\HP_{-1, -1, 0} + \frac{2}{3}\*\HP_{-1, 0, 0}\bigg]
+ (1 + x)\*\bigg[9\*\z3 + \frac{9}{2}\*\z2
 + 6\*\z2\*\HP_{0}
\nonumber\\&&\quad
 + 18\*\z2\*\HP_{-1}
 - \frac{13}{3}\*\HP_{0} - \frac{14}{3}\*\HP_{2} +
   12\*\HP_{-2, 0} - \frac{1}{3}\*\HP_{-1, 0} - 24\*\HP_{-1, 2} -
   12\*\HP_{-1, -1, 0}
 - 6\*\HP_{-1, 0, 0}
\nonumber\\&&\quad
 + 6\*\HP_{0, 0, 0}\bigg]
+ x^2\*\bigg[\frac{1}{3}\*\z2 + \frac{8}{3}\*\z3 + \frac{10}{3}\*\z2\*\HP_{0} +
   \frac{8}{3}\*\z2\*\HP_{1} - \frac{8}{3}\*\HP_{3} -
   \frac{4}{3}\*\HP_{-2, 0} - \frac{1}{3}\*\HP_{0, 0} -
   \frac{2}{3}\*\HP_{0, 0, 0}\bigg]
\nonumber\\&&\quad
+ \pqqmx\*\bigg[
     -7 \*\z2^2
     + 9\*\z3
     - 64\*\z2\*\HP_{-2}
     + \Big(-12\*\z2 - 72\*\z3\Big)\*\HP_{-1}
     + \Big(3 + 3\*\z2 + 26\*\z3\Big)\*\HP_{0}
\nonumber\\&&\quad
     - 12\*\z2\*\HP_{2}
     - 6\*\HP_{3}
     - 24\*\HP_{4}
     + 12\*\HP_{-3, 0}
     - 6\*\HP_{-2, 0}
     + 56\*\HP_{-2, 2}
     + 96\*\z2\*\HP_{-1, -1}
     - 80\*\z2\*\HP_{-1, 0}
\nonumber\\&&\quad
     + 12\*\HP_{-1, 2}
     + 64\*\HP_{-1, 3}
     + 28\*\z2\*\HP_{0, 0}
     - 4\*\HP_{3, 0}
     + 52\*\HP_{-2, 0, 0}
     - 6\*\HP_{-1, 0, 0}
     + 8\*\HP_{-1, 2, 0}
     + 9\*\HP_{0, 0, 0}
\nonumber\\&&\quad
     - 16\*\HP_{-2, -1, 0}
     - 16\*\HP_{-1, -2, 0}
     - 96\*\HP_{-1, -1, 2}
     - 80\*\HP_{-1, -1, 0, 0}
     + 44\*\HP_{-1, 0, 0, 0}
     - 12\*\HP_{0, 0, 0, 0}
     \bigg]
\bigg)
\nonumber\\&&\quad
+16\*\ca\*\cf\*\left(\cf-\frac{\ca}{2}\right)\*
\bigg(
       45\*\z3
     - 2\*\HP_{0}
     + 36\*\z2\*\HP_{0}
     - 18\*\HP_{3}
     + 18\*\HP_{-2, 0}
     + \HP_{0, 0}
\nonumber\\&&\quad
+ (1 - x)\*\bigg[
     - \frac{1}{2}\*\z2
     - 27\*\z3
     - 27\*\z2\*\HP_{0}
     +\frac{179}{18}
     - 9\*\z2\*\HP_{1}
     - \frac{17}{3}\*\HP_{1}
     + 18\*\HP_{3}
     - \HP_{0, 0}\bigg]
\nonumber\\&&\quad
+ (1 + x)\*\bigg[
     - 18\*\z3
     - \frac{3}{2}\*\z2
     - 9\*\z2\*\HP_{-1}
     + \frac{5}{3}\*\HP_{0}
     - 9\*\z2\*\HP_{0}
     + 2\*\HP_{2}
     - 18\*\HP_{-2, 0}
     + \HP_{-1, 0}
\nonumber\\&&\quad
     + 18\*\HP_{-1, 2}
     + 18\*\HP_{-1, -1, 0}\bigg]
+ x^2\*\bigg[
     - \z3
     + \frac{1}{3}\*\z2
     - 2\*\z2\*\HP_{0}
     - 2\*\z2\*\HP_{1}
     + 2\*\HP_{3}
     + 2\*\HP_{-2, 0}
\nonumber\\&&\quad
     - \frac{1}{3}\*\HP_{0, 0}\bigg]
+ \left(\frac{1}{x} + x^2\right)\*\bigg[
       \z2\*\HP_{-1}
     + \z2\*\HP_{1}
     + \frac{1}{3}\*\HP_{-1, 0}
     - 2\*\HP_{-1, 2}
     - 2\*\HP_{-1, -1, 0}\bigg]
\nonumber\\&&\quad
+ \pqqmx\*\bigg[
       \frac{134}{9}\*\z2
     - 4\*\z2^2
     - 11\*\z3
     + 32\*\z2\*\HP_{-2}
     + \left(
           48\*\z3
         + \frac{44}{3}\*\z2\right)\*\HP_{-1}
     + 8\*\z2\*\HP_{2}
     + \frac{22}{3}\*\HP_{3}
\nonumber\\&&\quad
     + 8\*\HP_{4}
     + \left(
         - 16\*\z3
         - \frac{2}{3}\*\z2
         - 3\right)\*\HP_{0}
     - 4\*\HP_{-3, 0}
     + \frac{22}{3}\*\HP_{-2, 0}
     - 32\*\HP_{-2, 2}
     - 64\*\z2\*\HP_{-1, -1}
\nonumber\\&&\quad
     + \left(
           44\*\z2
         + \frac{268}{9}\right)\*\HP_{-1, 0}
     - \frac{44}{3}\*\HP_{-1, 2}
     - 32\*\HP_{-1, 3}
     + \left(
         - 12\*\z2
         - \frac{134}{9}\right)\*\HP_{0, 0}
     - 16\*\HP_{-2, 0, 0}
\nonumber\\&&\quad
     + 64\*\HP_{-1, -1, 2}
     + \frac{22}{3}\*\HP_{-1, 0, 0}
     - \frac{31}{3}\*\HP_{0, 0, 0}
     + 32\*\HP_{-1, -1, 0, 0}
     - 12\*\HP_{-1, 0, 0, 0}
     + 4\*\HP_{0, 0, 0, 0}
     \bigg]
\bigg)\,.\label{eq:Pqq2m}
\end{eqnarray}

\setcounter{equation}{0}
\section{Conclusion}\label{sec:Conclusion}

We have found the complete third-order contributions to the splitting functions governing the evolution of flavour non-singlet transversity parton distribution in perturbative QCD. Firstly, we calculate the fifteen moments of the anomalous dimension of the non-singlet transversity operator with the help of MINCER package~\cite{Gorishnii:1989gt} for FORM~\cite{Vermaseren:2000nd}. As in our previous paper~\cite{Velizhanin:2010cm} for the reconstruction of the complete anomalous dimension for arbitrary moment $N$ from the know fixed values we have used \texttt{LLL}-algorithm~\cite{Lenstra:1982}, that allow to find coefficients in the ansatz from the harmonic sums, which form the suitable basis.
We have assumed, that the most complicated part of anomalous dimension could be obtained by replacement $\Nminus\to 1$ and $\Nplus\to 1$ in the known result for the non-singlet anomalous dimension from Ref.~\cite{Moch:2004pa}.
We used the same procedure in Ref.~\cite{Kotikov:2004er} together with the maximal transcendentality principle for the computation of the three-loop universal anomalous dimension in the maximally extended $\cN=4$ SYM theory. In present paper, we have subtracted this part from the known fifteen moments of anomalous dimension, listed in Appendix. The obtained numbers look much more simple with compare to the original numbers from Eqs.~(\ref{res1})-(\ref{res15})
and we can expect that the basis will reduced considerably. Apply \texttt{LLL}-algorithm to the minimal basis and the highest even or odd obtained numbers we have reconstructed the complete three-loop anomalous dimension of the flavour non-singlet transversity operator for arbitrary $N$.
From obtained expressions the \mbox{$x$-space} splitting functions can be obtained by a standard Mellin inversion with HARMPOL package~\cite{Vermaseren:1998uu,Remiddi:1999ew} for the symbolic manipulation program FORM~\cite{Vermaseren:2000nd} and we present the corresponding results in section~\ref{sec:ResultsX}.

For possible test of our result we give below an explicit expression for the three-loop anomalous dimension $\gamma_{\mathrm{TR},\mathrm{ns}}^{(2)}(\M)$ from Eq.~(\ref{eqgqq1p}) at $\M=16$:
\begin{eqnarray}
\gamma_{\mathrm{TR,ns}}^{(2)}(16)&=&
  - \frac{9588431966492629867}{3579902263121184000}\,\*\cf\*\nf^2
\nonumber\\[1mm]&&
  - \ca\*\cf\*\nf\*\left(\frac{26209369318544258455663}{1002372633673931520000}
     + \frac{3792038}{45045}\*\,\z3\right)
\nonumber\\[1mm]&&
  - \cf^2\*\nf\*\left(\frac{214170060836708111694659}{2810574247360239360000}
     - \frac{3792038}{45045}\*\,\z3\right)
\nonumber\\[1mm]&&
  - \cf^3\*\left(\frac{5187682484935485448445159707}{303845560733620756730880000}
     - \frac{100658462371}{2705402700}\*\,\z3\right)
\nonumber\\[1mm]&&
  - \ca\*\cf^2\*\left(\frac{24000075522445759616746817}{590220591945650265600000}
     + \frac{100658462371}{1803601800}\*\,\z3\right)
\nonumber\\[1mm]&&
  + \ca^2\*\cf\*\left(\frac{1448460151630990534353391}{5613286748574016512000}
     + \frac{100658462371}{5410805400}\*\,\z3\right).
\end{eqnarray}

The result of the recent calculations for $N=16$ Mellin moment of the three-loop anomalous dimension of the non-singlet transversity operator from Ref.~\cite{Bagaev:2012bw} coincides with the above prediction and can serve as a direct confirmation of the general result presented in this paper, at least for even values of $N$.

{ FORM} file of our results can be obtained from the preprint server \ {\tt http://arXiv.org} by downloading the source.
Furthermore it is available from author upon request.

\subsection*{Acknowledgments}
This work is supported by
RFBR grants 10-02-01338-a, 12-02-00412-a, RSGSS-65751.2010.2.

\newpage
\section*{Appendix: The 3--loop Anomalous Dimensions}\label{AppA}
\renewcommand{\theequation}{A.\arabic{equation}}
\setcounter{equation}{0}

\begin{eqnarray}
\gamma_{\mathrm{TR},\mathrm{ns}}^{(2)}(1)&=&
- \frac{4}{3}\, \cf\*\n2f\*\T2F
- \nf\*\TF\*\left(\ca\*\cf\*\left(\frac{1004}{27}+16\*\z3\right)
+ \cf^2\*\left(\frac{98}{9}+16\*\z3 \right)\right)\nonumber\\[3mm]&&\hspace*{-15mm}
+ \cf^3\*\left(\frac{365}{6}-64\*\z3 \right)
- \ca\*\cf^2\*\left(\frac{6823}{36}-112\*\z3\right)
+ \ca^2\*\cf\*\left(\frac{13639}{108}-40\*\z3\right)
,\label{res1}\\[6mm]
\gamma_{\mathrm{TR},\mathrm{ns}}^{(2)}(2)&=&
- \frac{92}{27}\,\cf\*\n2f\*\T2F
-\nf\*\TF\*\left(\cf^2\*\left(30 - 48\*\z3\right)
- \ca\*\cf\*\left(\frac{1042}{27}+48\*\z3 \right)\right)\nonumber\\[3mm]&&\hspace*{-10mm}
+ \frac{33}{2}\cf^3
- \frac{277}{4}\ca\*\cf^2
+ \frac{12553}{108}\ca^2\*\cf
,\label{res2}\\[6mm]
\gamma_{\mathrm{TR},\mathrm{ns}}^{(2)}(3)&=&
- \frac{1204}{243}\,\cf\*\n2f\*\T2F
- \nf\*\TF\*\left(\ca\*\cf\*\left(\frac{9725}{243}
+ \z3\*\frac{208}{3}\right)
+ \cf^2\*\left(\frac{12638}{243}
+ \z3\*\frac{208}{3}\right)\right)\nonumber\\[3mm]&&\hspace*{-12mm}
- \cf^3\*\left(\frac{8717}{486}
- \z3\*\frac{64}{3}\right)
- \ca\*\cf^2\*\left(\frac{30197}{972}+32\*\z3\right)
+ \ca^2\*\cf\*\left(
\frac{126557}{972}
+\z3\*\frac{32}{3}
\right),\label{res3}\\[6mm]
\gamma_{\mathrm{TR},\mathrm{ns}}^{(2)}(4)&=&
-\frac{7361}{1215}\,\cf\*\n2f\*\T2F
- \nf\*\TF\*\left(
\ca\*\cf\*\left(
\frac{199723}{4860}
+ \z3\*\frac{256}{3}
\right)
+ \cf^2\*\left(
\frac{66443}{972}
+ \z3\*\frac{256}{3}
\right)\right)\nonumber\\[3mm]&&\hspace*{-16mm}
- \cf^3\*\left(\frac{11755}{972}-28\*\z3\right)
-\ca\*\cf^2\*\left(\frac{132431}{3888}+42\*\z3\right)
+ \ca^2\*\cf\*\left(\frac{2893009}{19440}+14\*\z3 \right)
,\label{res4}\\[6mm]
\gamma_{\mathrm{TR},\mathrm{ns}}^{(2)}(5)&=&
-\frac{209297}{30375}\,\cf\*\n2f\*\T2F
- \nf\*\TF\*\ca\*\cf\*\left(\frac{5113951}{121500}
+ \z3\*\frac{1472}{15}\right)\nonumber\\[3mm]&&\hspace*{-10mm}
- \nf\*\TF\cf^2\*\left(\frac{49495163}{607500}
- \z3\*\frac{1472}{15}\right)
- \cf^3\*\left(\frac{15920231}{759375}
- \z3\*\frac{2356}{75}\right)\nonumber\\[3mm]&&\hspace*{-10mm}
- \ca\*\cf^2\*\left(\frac{65092847}{2430000}
+ \z3\*\frac{1178}{25}\right)
+ \ca^2\*\cf\*\left(
\frac{79012363}{486000}
+ \z3\*\frac{1178}{75}
\right)
,\label{res5}\\[6mm]
\gamma_{\mathrm{TR},\mathrm{ns}}^{(2)}(6)&=&
-\frac{1604879}{212625}\,\cf\*\n2f\*\T2F
- \nf\*\TF\*\ca\*\cf\*\left(\frac{1841332}{42525}
+ \z3\*\frac{544}{5}\right)\nonumber\\[3mm]&&\hspace*{-10mm}
- \nf\*\TF\cf^2\*\left(\frac{18622301}{202500}
- \z3\*\frac{544}{5}\right)
- \cf^3\*\left(\frac{4075862}{253125}
- \z3\*\frac{832}{25}\right)\nonumber\\[3mm]&&\hspace*{-10mm}
- \ca\*\cf^2\*\left(\frac{2180171}{67500}
+ \z3\*\frac{1248}{25}\right)
+ \ca^2\*\cf\*\left(
\frac{11152459}{63000}
+\z3\*\frac{416}{25}\right)
,\label{res6}\\[6mm]
\gamma_{\mathrm{TR},\mathrm{ns}}^{(2)}(7)&=&
-\frac{84250571}{10418625}\,\cf\*\n2f\*\T2F
- \nf\*\TF\*\ca\*\cf\*\left(\frac{1844723441}{41674500}
+ \z3\*\frac{4128}{35}\right)\nonumber\\[3mm]&&\hspace*{-10mm}
- \nf\*\TF\cf^2\*\left(\frac{49282560541}{486202500}
- \z3\*\frac{4128}{35}\right)
- \cf^3\*\left(\frac{232913441459}{11344725000}
- \z3\*\frac{42208}{1225}\right)\nonumber\\[3mm]&&\hspace*{-10mm}
- \ca\*\cf^2\*\left(\frac{18979072067}{648270000}
+ \z3\*\frac{63312}{1225}\right)
+ \ca^2\*\cf\*\left(
\frac{5219842927}{27783000}
+ \z3\*\frac{21104}{1225}
\right)
,\label{res7}\\[5mm]
\gamma_{\mathrm{TR},\mathrm{ns}}^{(2)}(8)&=&
- \frac{711801943}{83349000}\,\cf\*\n2f\*\T2F
- \nf\*\TF\*\ca\*\cf\*\left(\frac{6056338297}{133358400}
+ \z3\*\frac{4408}{35}\right)\nonumber\\[3mm]&&\hspace*{-12mm}
- \nf\*\TF\cf^2\*\left(\frac{849420853541}{7779240000}
- \z3\*\frac{4408}{35}\right)
- \cf^3\*\left(\frac{2324068794763}{136136700000}
- \z3\*\frac{43153}{1225}\right)\nonumber\\[3mm]&&\hspace*{-12mm}
- \ca\*\cf^2\*\left(\frac{350888781989}{10372320000}
+ \z3\*\frac{129459}{2450}\right)
+ \ca^2\*\cf\*\left(
 \frac{177184521133}{889056000}
+ \z3\*\frac{43153}{2450}
\right)\!,\label{res8}\\[5mm]
\gamma_{\mathrm{TR},\mathrm{ns}}^{(2)}(9)&=&
- \frac{20096458061}{2250423000}\,\cf\*\n2f\*\T2F
- \nf\*\TF\*\ca\*\cf\*\left(\frac{119131812533}{2571912000}
+ \z3\*\frac{41912}{315}\right)\nonumber\\[3mm]&&\hspace*{-10mm}
- \nf\*\TF\cf^2\*\left(\frac{24479706761047}{210039480000}
- \z3\*\frac{41912}{315}\right)\nonumber\\[3mm]&&\hspace*{-10mm}
- \cf^3\*\left(\frac{3936113801653709}{198487308600000}
- \z3\*\frac{1183051}{33075}\right)\nonumber\\[3mm]&&\hspace*{-10mm}
- \ca\*\cf^2\*\left(\frac{9010105083551}{280052640000}
+ \z3\*\frac{1183051}{22050}\right)\nonumber\\[3mm]&&\hspace*{-10mm}
+ \ca^2\*\cf\*\left(
\frac{14990912737013}{72013536000}
+\z3\*\frac{1183051}{66150}
\right)
,\label{res9}\\[5mm]
\gamma_{\mathrm{TR},\mathrm{ns}}^{(2)}(10)&=&
-\frac{229508848783}{24754653000}\, \cf\*\n2f\*\T2F
- \nf\*\TF\*\ca\*\cf\*\left(\frac{4264058299021}{90016920000}
+ \z3\*\frac{43928}{315}\right)\nonumber\\[3mm]&&\hspace*{-10mm}
- \nf\*\TF\cf^2\*\left(\frac{25800817445759}{210039480000}
- \z3\*\frac{43928}{315}\right)\nonumber\\[3mm]&&\hspace*{-10mm}
- \cf^3\*\left(\frac{686447082796333}{39697461720000}
- \z3\*\frac{1195987}{33075}\right)\nonumber\\[3mm]&&\hspace*{-10mm}
- \ca\*\cf^2\*\left(\frac{12524164039049}{350065800000}
+ \z3\*\frac{1195987}{22050}\right)\nonumber\\[3mm]&&\hspace*{-10mm}
+ \ca^2\*\cf\*\left(
\frac{391644487915601}{1800338400000}
+ \z3\*\frac{1195987}{66150}\right)
,\label{res10}\\[4mm]
\gamma_{\mathrm{TR},\mathrm{ns}}^{(2)}(11)&=&
- \frac{28677274464343}{2995313013000}\cf\*\n2f\*\T2F\nonumber\\[2mm]&&\hspace*{-10mm}
- \nf\*\TF\*\ca\*\cf\*\left(\frac{75010870835743}{1556006760000}
+ \z3\*\frac{503368}{3465}\right)\nonumber\\[2mm]&&\hspace*{-10mm}
- \nf\*\TF\cf^2\*\left(\frac{396383896707569599}{3075188026680000}
- \z3\*\frac{503368}{3465}\right)\nonumber\\[2mm]&&\hspace*{-10mm}
- \cf^3\*\left(\frac{12285093670097281643}{639331590746772000}
- \z3\*\frac{145890427}{4002075}\right)\nonumber\\[2mm]&&\hspace*{-10mm}
- \ca\*\cf^2\*\left(\frac{16180046111662399}{465937579800000}
+ \z3\*\frac{145890427}{2668050}\right)\nonumber\\[2mm]&&\hspace*{-10mm}
+ \ca^2\*\cf\*\left(
\frac{49020505864347881}{217840946400000}
+ \z3\*\frac{145890427}{8004150}
\right)
,\label{res11}\\[4mm]
\gamma_{\mathrm{TR},\mathrm{ns}}^{(2)}(12)&=&
- \frac{383379490933459}{38939069169000}\cf\*\n2f\*\T2F\nonumber\\[2mm]&&\hspace*{-10mm}
- \nf\*\TF\*\ca\*\cf\*\left(\frac{38283693844132279}{778781383380000}
+ \z3\*\frac{521848}{3465}\right)\nonumber\\[2mm]&&\hspace*{-10mm}
- \nf\*\TF \cf^2\*\left(\frac{1237841854306528417}{9225564080040000}
- \z3\*\frac{521848}{3465}\right)\nonumber\\[2mm]&&\hspace*{-10mm}
- \cf^3\*\left(\frac{88375433743256891029}{5114652725974176000}
- \z3\*\frac{146791327}{4002075}\right)\nonumber\\[2mm]&&\hspace*{-10mm}
- \ca\*\cf^2\*\left(\frac{6939941146982213189}{184511281600800000}
+ \z3\*\frac{146791327}{2668050}\right)\nonumber\\[2mm]&&\hspace*{-10mm}
+ \ca^2\*\cf\*\left(
\frac{1209479407264968533}{5191875889200000}
+ \z3\*\frac{146791327}{8004150}
\right)
,\label{res12}\\[4mm]
\gamma_{\mathrm{TR},\mathrm{ns}}^{(2)}(13)&=&
- \frac{66409807459266571}{6580702689561000}\,\cf\*\n2f\*\T2F\nonumber\\[2mm]&&\hspace*{-10mm}
- \nf\*\TF\*\ca\*\cf\*\left(\frac{6571493644375020121}{131614053791220000}
+ \z3\*\frac{7005784}{45045}\right)\nonumber\\[2mm]&&\hspace*{-10mm}
- \nf\*\TF\*\cf^2\*\left(\frac{36713319015407141570017}{263491335690022440000}
- \z3\*\frac{7005784}{45045}\right)\nonumber\\[2mm]&&\hspace*{-10mm}
- \cf^3\*\left(\frac{35539078759302008520407749}{1899034754585129729568000}
- \z3\*\frac{24927484663}{676350675}\right)\nonumber\\[2mm]&&\hspace*{-10mm}
- \ca\*\cf^2\*\left(\frac{64789158904559342386423}{1756608904600149600000}
+ \z3\*\frac{24927484663}{450900450}\right)\nonumber\\[2mm]&&\hspace*{-10mm}
+ \ca^2\*\cf\*\left(
\frac{70028470572016843639}{292475675091600000}
+ \z3\*\frac{24927484663}{1352701350}\right)
,\label{res13}\\[4mm]
\gamma_{\mathrm{TR},\mathrm{ns}}^{(2)}(14)&=&
- \frac{67887731189063371}{6580702689561000}\,\cf\*\n2f\*\T2F\nonumber\\[2mm]&&\hspace*{-10mm}
- \nf\*\TF\*\ca\*\cf\*\left(\frac{1063558335824080583}{20938599466785000}
+ \z3\*\frac{7211704}{45045}\right)\nonumber\\[2mm]&&\hspace*{-10mm}
- \nf\*\TF\*\cf^2\*\left(\frac{37914105069781083851977}{263491335690022440000}
- \z3\*\frac{7211704}{45045}\right)\nonumber\\[2mm]&&\hspace*{-10mm}
- \cf^3\*\left(\frac{163200953601404076174043511}{9495173772925648647840000}
- \z3\*\frac{25023044413}{676350675}\right)\nonumber\\[2mm]&&\hspace*{-10mm}
- \ca\*\cf^2\*\left(\frac{131593571838833571500893}{3353526090600285600000}
+ \z3\*\frac{25023044413}{450900450}\right)\nonumber\\[2mm]&&\hspace*{-10mm}
+ \ca^2\*\cf\*\left(
\frac{1323640493375923341991}{5374240529808150000}
+ \z3\*\frac{25023044413}{1352701350}
\right)
,\label{res14}\\[5mm]
\gamma_{\mathrm{TR},\mathrm{ns}}^{(2)}(15)&=&
- \frac{69246489611060347}{6580702689561000} \,\cf\*\n2f\*\T2F\nonumber\\[2mm]&&\hspace*{-10mm}
- \nf\*\TF\*\ca\*\cf\*\left(\frac{2965759870308771337}{57581148533658750}
+ \z3\*\frac{7403896}{45045}\right)\nonumber\\[2mm]&&\hspace*{-10mm}
+ \nf\*\TF\*\cf^2\*\left(\frac{13032900994209180547219}{87830445230007480000}
- \z3\*\frac{7403896}{45045}\right)\nonumber\\[2mm]&&\hspace*{-10mm}
- \cf^3\*\left(\frac{173845824913967604253242671}{9495173772925648647840000}
- \z3\*\frac{25100738029}{676350675}\right) \nonumber\\[2mm]&&\hspace*{-10mm}
- \ca\*\cf^2\*\left(\frac{1427678608519179211666397}{36888786996603141600000}
+ \z3\*\frac{25100738029}{450900450}\right)\nonumber\\[2mm]&&\hspace*{-10mm}
+ \ca^2\*\cf\*\left(\frac{325033784336407730099}{1289817727153956000}
+\z3\*\frac{25100738029}{1352701350}\right)
.\label{res15}
\end{eqnarray}

\end{document}